# Marangoni instability in a viscoelastic binary film with cross-diffusive effect

**Rajkumar Sarma** and **Pranab Kumar Mondal**†

Department of Mechanical Engineering, Indian Institute of Technology Guwahati, Assam, India – 781039

The viscoelastic fluids are usually the blends of a polymeric solute and a Newtonian solvent. In the presence of a temperature gradient, stratification of these solutes can take place via the Soret effect. Here, we investigate the classical Marangoni instability problem for a thin viscoelastic film considering this binary aspect of the fluid. The film, bounded above by a deformable free surface, is subjected to heating from below by a solid substrate. Linear stability analysis performed numerically for perturbations of finite wavelength (short-wave perturbations) reveals that both monotonic and oscillatory instabilities can emerge in this system. The interaction between the thermocapillary and solutocapillary forces in the presence of Soret diffusion is found to give rise to two different oscillatory instabilities, of which one mode was overlooked previously, even for the Newtonian binary mixtures. As a principal result of this work, we provide a complete picture of the susceptibility to different instability modes based on the model parameter values. Finally, an approximate model is developed under the framework of long-wave analysis, which can qualitatively depict the stability behaviour of the system without numerically solving the problem.

## 1. Introduction

A gradient of surface tension caused by the inhomogeneities in temperature (thermo-capillarity) or concentration (solutocapillarity) on the free surface of a pure liquid or liquid mixtures has the ability to induce motion in its bulk phase. Typically known as the Marangoni convection, this phenomenon is frequently encountered in the small-scale systems (e.g. a thin liquid film, droplet, vapor bubble, or a liquid bridge) where the surface effects dominate over the volumetric ones. Understanding the dynamics of Marangoni convection is essential for optimizing the operations like interfacial heat and mass transport (applications include the thin-film evaporation, liquid-liquid extraction), and the materials processing problems (semiconductor crystal growth, weld-deposition), especially for a microgravity environment where the buoyancy-driven Rayleigh-Bénard convection gets inhibited.

Since the founding experiments of Bénard (1901), several theoretical, experimental, and numerical investigations have been carried out over the years to elucidate the major features of this convection process. For a short historical account on these works, the reader is referred to the monographs and review papers by Oron *et al.* (1997), Colinet, Legros & Velarde (2001), Nepomnyashchy (2001), Schatz & Neitzel (2001), Craster & Matar (2009) and Shklyaev & Nepomnyashchy (2017). It is now well known that unlike the pure liquids, Marangoni convection in a multicomponent liquid film can develop under the simultaneous actions of thermocapillary and solutocapillary effects. Such liquid mixtures usually exhibit a strong Soret effect (a cross-diffusive effect that leads to spontaneous development of the solute concentration gradient under an imposed temperature gradient). Hence, two different physical situations are possible while investigating the Marangoni convection in a multicomponent liquid film: (i) the temperature and the concentration gradients are caused by independent sources (often called as the double-diffusive convection); and (ii) the temperature gradient is externally imposed, while the Soret effect generates the concentration gradient.

Motivated by their importance in materials processing, species separation in food, chemistry, and biomedical applications, both the aforementioned cases have been extensively studied in the literature. The double-diffusive problem was pioneered by McTaggert (1983) to analyze the linear stability characteristics of a horizontally infinite binary liquid film. Later, Ho & Chang (1988) extended this analysis to study the nonlinear aspect of the convection process; Arafune, Yamatoto & Hirata (2001) experimentally investigated this problem; while Chen, Li & Zhan (2010) tackled the problem for a confined cavity. More recently, D'Alessio *et al.* (2020) have theoretically analyzed the double-diffusive problem for a liquid film falling down a heated inclined plate. These analyses demonstrate that both monotonic and oscillatory instabilities are possible in a binary liquid film. The monotonic mode appears when the shear stresses induced by the thermal and solutal components enhance each other, and whenever they counteract, the oscillatory convection develops.

The cross-diffusive Marangoni convection problem has also been the subject of numerous investigations, starting with the precursor works of Bhattacharjee (1994), Joo (1995), and Skarda, Jacqmin & McCaughan (1998). These authors addressed the problem for a horizontal liquid film resting on an ideally thermally conductive substrate (i.e. considering a constant temperature bottom boundary condition) and studied the onset of instability under the framework of linear stability analysis. It was shown that, although the monotonic disturbances can emerge in the long-wave form in a non-deformable surface, the deformability of the free surface is essential for the appearance of long-wave oscillatory perturbations. For a poorly conductive substrate (i.e. the condition of fixed heat flux at the bottom boundary), later Oron & Nepomnyashchy (2004) detected a different kind of long-wave oscillatory disturbance that can develop even without surface deformations. Shklyaev, Nepomnyashchy & Oron (2009) then extended this analysis to decipher the short-wave mode of this particular oscillatory instability. Recent research in the field of cross-diffusive Marangoni convection has focused on exploring the role of free surface deformability (Podolny, Oron & Nepomnyashchy 2005; Hu *et al.* 2008; Bestehorn & Borcia 2010), the effect of surfactant adsorption/desorption on the free surface (Shklyaev & Nepomnyashchy 2013) or the influence of modulated boundary conditions (Fayzrakhmanova, Shklyaev & Nepomnyashchy 2013) on the stability behaviour of the system. However, it should be noted that all the above-mentioned works deal with a Newtonian binary mixture.

Despite such remarkable advancements towards understanding the Marangoni convection in Newtonian fluids, relatively little attention has been devoted to the viscoelastic fluids. Viscoelastic fluids, e.g. the polymeric solutions, biofluids, paints, lubricants, etc. exhibit complex rheological behaviour due to both the viscous and elastic character (Bird *et al.* 1987). A non-trivial relaxation time (a measure of the elasticity of the fluid) significantly alters the dynamics of such fluids from their Newtonian counterpart. Marangoni convection is often encountered in viscoelastic fluids during the phenomenon like the drying of a thin polymeric film (Toussaint *et al.* 2008; Bassou & Rharbi 2009; Bormashenko *et al.* 2010). The convective patters developed in such liquid films have promising properties for the nanotechnological applications, e.g. organic photovoltaics and photodiodes (Heriot & Jones 2005). It is important to note that viscoelastic fluids are usually composed of a polymeric solute and a Newtonian solvent, and hence, essentially a binary mixture. The Soret effect can yield a stratification of the solutes in such fluids as well (de Gans *et al.* 2003). Typically, while these solutes tend to migrate towards a colder region (owing to their large masses), sometimes, depending on the solvent quality and the temperature of the mixture, they can also move into the warmer region (Zhang & Müller-Plathe 2006; Würger 2007). Such migration of the solutes can lead to the development of solutocapillary stress on the free surface of a viscoelastic liquid film. This aspect necessitates the consideration of a complete thermosolutal model to investigate the Marangoni convection problem in a viscoelastic liquid.

In the previously reported studies on Marangoni instability in a viscoelastic film, this binary aspect of the fluid was either completely ignored; or the problem was analyzed separately for the thermal and solutal convections (i.e. without considering a complete thermosolutal model). A purely thermal model (Dauby *et al.* 1993; Getachew & Rosenblat 1985; Parmentier *et al.* 2000; Sarma & Mondal 2019) suggests the emergence of both monotonic and oscillatory disturbances in the system. The monotonic mode is found to get dominant in a weakly viscoelastic liquid, while, the oscillatory mode prevails in highly viscoelastic liquids. Nevertheless, such a model is inadequate to illustrate the instability modes caused by the solutocapillary force. On the other hand, the solutal problem was analyzed by Doumenc *et al.* (2013) and Yiantsios *et al.* (2015) in the context of evaporation in a polymeric film. In these works, the concentration gradient was considered to be solely caused by the difference in the evaporation rate between the constituents, neglecting the Soret effect. These analyses provide a deep insight into the problem regarding the onset of convection in the film and the evolution of the disturbances in the nonlinear regimes. However, the role of elasticity of the fluid on the film dynamics is not clear from these works, since the polymeric solution was treated as a Newtonian fluid. Furthermore, it also needs to be pointed out that separate thermal and solutal models are incapable of depicting the instability modes that may emerge from the interaction between them.

The present work aims at developing a complete thermosolutal model to investigate the Marangoni convection problem for a thin viscoelastic film. The fluids considered here can spontaneously generate a concentration gradient via the Soret effect on the imposition of a temperature gradient, e.g. poly(ethylene oxide)/water, poly(vinyl alcohol)/water, polystyrene/dioctyl phthalate (Zhang & Müller-Plathe 2006). Employing a viscoelastic constitutive model to depict the rheology of the fluid, we study the stability characteristics of the system under the framework of linear analysis. Besides exploring the role of fluid elasticity on the underlying convection, the present analysis also demonstrates the instability modes originating from the interaction between the thermocapillary and solutocapillary forces. In light of the results obtained in this work, another principal goal of this paper is to encourage future work with viscoelastic fluids.

The remainder of the paper proceeds as follows: in § 2, we formulate the problem by presenting the set of governing equations and boundary conditions. Linear stability analysis of the system is then carried out in § 3. The stability picture generated by numerically solving the eigenvalue problem is analyzed in § 4. In § 5, we study the effect of elasticity on the spatial structure of the eigenvectors at the neutral stability point. An approximate model is then introduced in § 6. We plot the phase diagrams in § 7, to provide a comprehensive picture of the susceptibility to different instability modes based on the model parameter values. And finally, the conclusions are drawn in § 8.

## 2. Mathematical model

### 2.1. *Problem statement and the governing equations*

We begin by considering a thin, two-dimensional layer of an incompressible viscoelastic polymer solution in the gravitational field **g** (see figure 1). The solution is a binary mixture of a polymeric solute and a Newtonian solvent, characterized by the relaxation time $\lambda$, viscosity $\mu_o (= \mu_s + \mu_p, \mu_s$ and $\mu_p$ are the solvent and solute viscosity, respectively), density $\rho$, thermal conductivity $\kappa$, thermal diffusivity $\alpha$, mass diffusivity $D$, and surface tension $\sigma$.

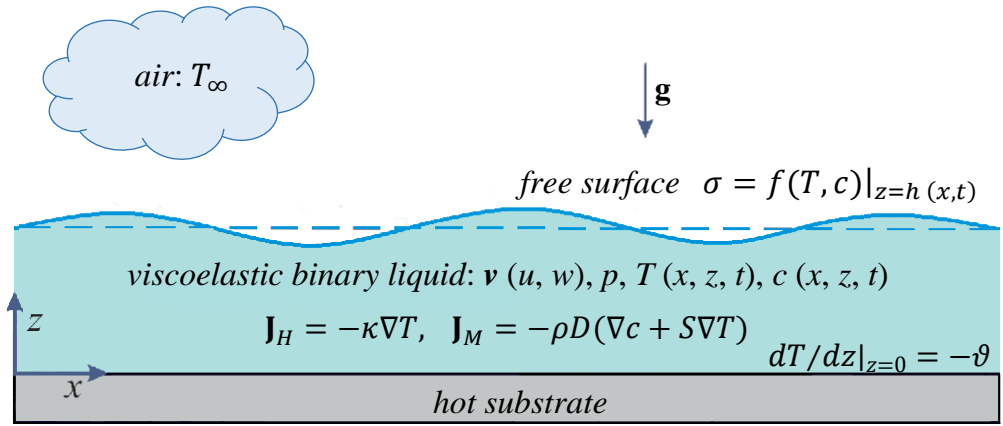

FIGURE 1. (Color online) Schematic illustration of the physical system under consideration. A thin viscoelastic film (composed of a polymeric solute in a Newtonian solvent), confined between its deformable free surface $z = h(x,t)$ and a horizontal substrate in the gravitational field **g**, is subjected to a vertical temperature gradient. This applied temperature gradient induces a concentration gradient in the film via the Soret effect. The surface tension gradient, arising from the inhomogeneities in temperature and concentration at the air-liquid interface, induces Marangoni convection in the fluid layer. The dashed line corresponds to the undeformed interface at the quiescent base state.

We consider the film to be of the infinite horizontal extent $x \in (-\infty, \infty)$ with unperturbed thickness $H$. At the $z = 0$ plane, the film is in thermal contact with a poorly conductive rigid substrate, and a deformable free surface located at $z = h(x,t)$ separates the film from the ambient gas phase. A transverse temperature gradient exists in the entire binary mixture, which is specified to be $-\vartheta$ at $z = 0$. This signifies $\vartheta > 0$ ($\vartheta < 0$) corresponds to the case of heating the fluid layer at the substrate (gas) side. The incorporation of the Soret effect into the analysis indicates that mass flux in the flow domain is a combination of the concentration and temperature gradients (Groot & Mazur 2011). Hence, the heat $\mathbf{J}_H$ and mass $\mathbf{J}_M$ fluxes within the film are governed by,

$$\mathbf{J}_H = -\kappa \nabla T, \qquad (2.1a)$$

$$\mathbf{J}_M = -\rho D (\nabla c + \mathcal{S} \nabla T), \qquad (2.1b)$$

where $\mathcal{S}$ is the Soret coefficient of the mixture. For a polymeric solution, $\mathcal{S}$ can be either positive or negative depending on the solvent quality, the mole fractions of the components, and the temperature of the binary mixture (Zhang & Müller-Plathe 2006). Note that a negative (positive) sign of $\mathcal{S}$ signifies a migration of the polymeric solutes towards the warmer (colder) region. It follows from (2.1) that, at the conductive state, the externally applied heat flux generates a temperature difference $\Delta T = \vartheta H$, which, in turn, yields a concentration difference $\Delta c = -\mathcal{S} \Delta T$ across the layer.

Now, above a particular temperature gradient, the thermo- and solutocapillary effects induce Marangoni convection in this mixture. The buoyancy effect is neglected in this study considering the small thickness of the film ($H \lesssim O(1)$ cm, see Pearson 1958). We assume the surface tension to vary linearly with temperature and solute concentration, dictated by the relationship:

$$\sigma = \sigma_o - \sigma_T (T - T_o) + \sigma_c (c - c_o), \qquad (2.2)$$

where $\sigma_o$ is the surface tension at the reference temperature $T_o$ and concentration $c_o$; $\sigma_T = -d\sigma/dT$ and $\sigma_c = d\sigma/dc$ quantifies the rate of change of surface tension with respect to the temperature and concentration. It should be noted that for most of the polymeric solutions, $(\sigma_T, \sigma_c) > 0$ (Doumenc et al. 2013). However, except $\sigma$, all other thermophysical properties are assumed to remain invariant of temperature in this analysis.

In the presence of linear Soret effect, the equations governing the fields of fluid velocity

$v \equiv \{u(x,z,t), w(x,z,t)\}$, pressure $p(x,z,t)$, temperature $T(x,z,t)$ and concentration $c(x,z,t)$ in the bulk of the film are given by,

$$\nabla \cdot v = 0, \qquad (2.3a)$$

$$\rho\left(\frac{\partial v}{\partial t} + v \cdot \nabla v\right) = -\nabla p + \nabla \cdot \tau - \rho g\, k, \qquad (2.3b)$$

$$\frac{\partial T}{\partial t} + v \cdot \nabla T = \alpha\, \nabla^2 T, \qquad (2.3c)$$

$$\frac{\partial c}{\partial t} + v \cdot \nabla c = D\nabla^2 c + \mathcal{S}D\, \nabla^2 T, \qquad (2.3d)$$

respectively, where $t$ represents time, $\tau = \begin{bmatrix} \tau_{xx} & \tau_{xz} \\ \tau_{zx} & \tau_{zz} \end{bmatrix}$ is the deviatoric stress tensor, $k$ is the unit vector in the $z$-direction, and $\nabla \equiv \{\partial/\partial x, \partial/\partial z\}$. The above set of governing equations are accompanied by the following boundary conditions:

At the $z = 0$ plane, where the film is in thermal contact with a rigid substrate, the pertinent boundary conditions encompass the no-slip, no penetration condition for velocity, a specified uniform normal heat flux, and the mass impermeability condition, represent respectively by,

$$v = 0, \quad \frac{\partial T}{\partial z} = -\vartheta, \quad \frac{\partial c}{\partial z} = \mathcal{S}\vartheta \qquad \text{at } z = 0. \qquad (2.4a\text{-}c)$$

At the deformable free surface $z = h(x,t)$, the boundary conditions comprise of the kinematic boundary condition, heat exchange with the ambient-gas phase (characterized by Newton's law of cooling), mass impermeability, and the balance of tangential and normal stress components, represented respectively by (2.5a-e):

$$w = \frac{\partial h}{\partial t} + u\, \frac{\partial h}{\partial x}, \qquad (2.5a)$$

$$-\kappa\left(\frac{\partial h}{\partial x}\frac{\partial T}{\partial x} - \frac{\partial T}{\partial z}\right) + q(T - T_\infty)\sqrt{1 + (\partial h/\partial x)^2} = 0, \qquad (2.5b)$$

$$\kappa\left(-\frac{\partial h}{\partial x}\frac{\partial c}{\partial x} + \frac{\partial c}{\partial z}\right) - \mathcal{S}q(T - T_\infty)\sqrt{1 + (\partial h/\partial x)^2} = 0, \qquad (2.5c)$$

$$\frac{1}{\sqrt{1 + (\partial h/\partial x)^2}}\left\{\tau_{xz}\left[1 - \left(\frac{\partial h}{\partial x}\right)^2\right] + \tau_{zz}\frac{\partial h}{\partial x} - \tau_{xx}\frac{\partial h}{\partial x}\right\} = \frac{\partial \sigma}{\partial x} + \frac{\partial \sigma}{\partial z}\frac{\partial h}{\partial x}, \qquad (2.5d)$$

$$-p + \frac{1}{1 + (\partial h/\partial x)^2}\left[\tau_{zz} + \tau_{xx}\left(\frac{\partial h}{\partial x}\right)^2 - 2\,\tau_{xz}\frac{\partial h}{\partial x}\right] = \sigma\,\frac{\partial^2 h/\partial x^2}{\left[1 + (\partial h/\partial x)^2\right]^{3/2}}$$

$$\text{at } z = h(x,t), \quad (2.5e)$$

In (2.5*b*,*c*), *q* denotes the rate of heat exchange between the free surface and the ambient air at temperature $T_\infty$. The kinematic boundary condition (2.5*a*) gives the location of the interface, while the mass impermeability condition (2.5*c*) portrays the non-volatile behaviour of the binary mixture. The dynamics of the gas-phase are decoupled here from the liquid phase considering large differences in the physical properties between both the phases.

## 2.2. *Constitutive model for the fluid*

Viscoelastic fluids exhibit complex rheology under the simultaneous action of the viscous and elastic character. Unlike the Newtonian fluids, the stress exhibits here an elastic response to the strain characterized by the relaxation time of the fluid. A wide variety of constitutive relationships, comprising of both the linear and nonlinear models, have been developed over the years to describe the rheology of viscoelastic fluids. In this analysis, we proceed with the linear Maxwell model (Maxwell 1867):

$$\boldsymbol{\tau} + \lambda \frac{\partial \boldsymbol{\tau}}{\partial t} = \mu_o \left[ (\nabla \boldsymbol{v}) + (\nabla \boldsymbol{v})^T \right], \quad (2.6)$$

which characterizes the fluid by a single relaxation time $\lambda$ [†] without incorporating the rheological nonlinearity. The reasons for adopting this particular constitutive model for this investigation are as follows: First, in the present convection phenomenon, motion is developed in a liquid film which was at rest in its equilibrium state. This indicates the shear rates involved with the underlying process are also extremely small. A nonlinear model (*viz.* the upper-convected Maxwell model, wherein the ordinary time derivative of $\boldsymbol{\tau}$ in (2.6) is replaced by the "upper-convected" time derivatives), is essential to describe the flow dynamics only at high shear rates. Second, since a linear stability analysis will be carried out around a quiescent base state, any nonlinear terms in the constitutive equation will not make here any contribution to the final linearized set of equations. The stability picture obtained using a linear model will be identical to that with the inclusion of the upper-convected terms. The aspects mentioned above suggest that the linearized Maxwell model is deemed sufficient to reveal the basic effect of elasticity for this analysis.

---

[†] $\lambda$ is interpreted here as the longest relaxation time out of the spectrum of relaxation time exhibited by a viscoelastic fluid.

## 2.3. Non-dimensionalization

The boundary value problem (BVP) formulated by (2.3)−(2.5) is now non-dimensionalized considering the unperturbed film thickness $H$ as the characteristic length scale, the thermal diffusion time $H^2/\alpha$ as the characteristic time scale, and $\vartheta H$ as the temperature scale.

This allows us to define the following set of dimensionless variables:

$$\left.\begin{array}{c} (\bar{x},\bar{z}) = \dfrac{(x,z)}{H}, \quad \bar{h} = \dfrac{h}{H}, \quad \bar{t} = \dfrac{t}{H^2/\alpha}, \quad (\bar{u},\bar{w}) = \dfrac{u,w}{(\alpha/H)}, \quad \bar{\tau} = \dfrac{\tau}{\mu\alpha/H^2}, \\ \bar{p} = \dfrac{p}{\mu\alpha/H^2}, \quad \bar{T} = \dfrac{T-T_\infty}{\vartheta H}, \quad \bar{c} = \dfrac{c}{\sigma_T \vartheta H / \sigma_c}. \end{array}\right\} \quad (2.7)$$

Note that, although the bulk concentration $c$ (defined either as the mass fraction or volume fraction) is a dimensionless quantity, its rescaling in the above-mentioned manner keeps this analysis coherent with the previously reported studies (Podolny *et al.* 2005; Shklyaev *et al.* 2009; Sarma & Mondal 2018). With this choice of the non-dimensional variables, we finally obtain the governing equations and the boundary conditions (dropping the overbar sign for notational convenience) in the following dimensionless form:

$$\nabla \cdot \boldsymbol{v} = 0, \tag{2.8a}$$

$$Pr^{-1}\left(\frac{\partial \boldsymbol{v}}{\partial t} + \boldsymbol{v}\cdot\nabla\boldsymbol{v}\right) = -\nabla p + \nabla \cdot \boldsymbol{\tau} - Ga\,\boldsymbol{k}, \tag{2.8b}$$

$$\frac{\partial T}{\partial t} + \boldsymbol{v}\cdot\nabla T = \nabla^2 T, \tag{2.8c}$$

$$\frac{\partial c}{\partial t} + \boldsymbol{v}\cdot\nabla c = Le\left(\nabla^2 c + \chi \nabla^2 T\right); \tag{2.8d}$$

$$\boldsymbol{v} = \boldsymbol{0}, \quad \frac{\partial T}{\partial z} = -1, \quad \frac{\partial c}{\partial z} = \chi \qquad \text{at } z = 0, \tag{2.9a-c}$$

$$w = \frac{\partial h}{\partial t} + u\frac{\partial h}{\partial x}, \tag{2.10a}$$

$$\left(\frac{\partial T}{\partial z} - \frac{\partial h}{\partial x}\frac{\partial T}{\partial x}\right) + Bi\,T\sqrt{1+(\partial h/\partial x)^2} = 0, \tag{2.10b}$$

$$\left(\frac{\partial c}{\partial z} - \frac{\partial h}{\partial x}\frac{\partial c}{\partial x}\right) - \chi Bi\,T\sqrt{1+(\partial h/\partial x)^2} = 0, \tag{2.10c}$$

$$-p + \frac{1}{1+(\partial h/\partial x)^2}\left[\tau_{zz} + \tau_{xx}\left(\frac{\partial h}{\partial x}\right)^2 - 2\tau_{xz}\frac{\partial h}{\partial x}\right] = \Sigma \frac{\partial^2 h/\partial x^2}{\left[1+(\partial h/\partial x)^2\right]^{3/2}}, \quad (2.10d)$$

$$\frac{1}{\sqrt{1+(\partial h/\partial x)^2}}\left\{\tau_{xz}\left[1-\left(\frac{\partial h}{\partial x}\right)^2\right] + \tau_{zz}\frac{\partial h}{\partial x} - \tau_{xx}\frac{\partial h}{\partial x}\right\} = Ma\left[-\frac{\partial T}{\partial x} + \frac{\partial c}{\partial x} + \left(-\frac{\partial T}{\partial z} + \frac{\partial c}{\partial z}\right)\frac{\partial h}{\partial x}\right]$$
$$\text{at } z = h(x,t). \quad (2.10e)$$

Moreover, in non-dimensional form, the Maxwell constitutive model (2.6) reads

$$\boldsymbol{\tau} + De\frac{\partial \boldsymbol{\tau}}{\partial t} = \left[(\nabla \boldsymbol{v}) + (\nabla \boldsymbol{v})^T\right]. \quad (2.11)$$

This problem is now characterized by the following set of dimensionless parameters: the Marangoni number, $Ma$, the Prandtl number, $Pr$, the Deborah number, $De$, the (inverse) Lewis number, $Le$, the Soret number, $\chi$, the Biot number, $Bi$, the Galileo number, $Ga$, and the (inverse) capillary number, $\Sigma$:

$$\left.\begin{array}{l} Ma = \dfrac{\sigma_T \vartheta H^2}{\mu_o \alpha}, \quad Pr = \dfrac{\mu_o}{\rho \alpha}, \quad De = \dfrac{\lambda \alpha}{H^2}, \quad Le = \dfrac{D}{\alpha}, \\[2mm] \chi = \dfrac{\mathcal{S}\sigma_c}{\sigma_T}, \quad Bi = \dfrac{qH}{\kappa}, \quad Ga = \dfrac{\rho g H^3}{\mu_o \alpha}, \quad \Sigma = \dfrac{\sigma H}{\mu_o \alpha}. \end{array}\right\} \quad (2.12)$$

The Marangoni number governs the present instability phenomenon. For the convection to set in, the thermo-solutocapillary stresses must have to overcome the stabilization effects of viscous and thermal diffusion. $Ma$ gives the critical temperature difference across the film ($\vartheta H$) at which the convection appears in the film. Note that, for $\sigma_T > 0$, $Ma$ can assume both positive and negative values depending on the direction of the applied temperature gradient $\vartheta$. A positive (negative) $Ma$ indicates the fluid layer is subjected to heating from below (above). We restrict this analysis only to the positive values of $Ma$. The Prandtl number is a material property of the fluid. Excluding the rarefied gases (which display a strong viscoelastic character with $Pr \ll 1$), for most of the viscoelastic fluids $Pr \gg 1$. This indicates a larger thermal diffusion time scale ($H^2/\alpha$) compared to the viscous diffusion time scale ($\rho H^2/\mu_o$) for such fluids. The Deborah number is a measure of the elasticity of the fluid. $De = 0$ indicates a Newtonian liquid ($\lambda = 0$), while higher values of $De$ signifies enhanced elastic behaviour of the fluid. The (inverse) Lewis number compares the characteristic mass diffusion time scale $H^2/D$ to the thermal diffusion time scale $H^2/\alpha$. $Le$ is usually small for a binary liquid mixture and lies within the range $10^{-5} \leq Le \leq 10^{-1}$. The Soret number takes into account the relative contribution of the thermocapillary and solutocapillary force to the free surface force. Note that, $\chi$ can be either positive or negative based on the Soret coefficient $\mathcal{S}$ (see § 2.1). The typical value of $\chi$ varies in the range $-1 \leq \chi \leq 1$. The Biot number characterizes the heat transfer rate across the free surface. The Galileo number and the (inverse) capillary number takes account of the deformability of the free surface through the magnitude of $g$ and $\sigma$. For a 0.1 mm thick layer of a polymeric solution with $\mu_o = O(10^{-2})$ Pas, $\rho = O(10^3)$ kg/m$^3$, $\alpha = O(10^{-7})$ m$^2$/s, $\sigma = O(10^{-2})$ N/m, and $g = O(0.1)$ m/s$^2$, we obtain $Ga = 0.1$ which corresponds to the microgravity environment. It is important to note that a free surface can be treated as non-deformable in the limit $(Ga, \Sigma) \to \infty$, which is usually the case of a liquid layer with very high surface tension placed at the terrestrial environment. In order to reveal the role of surface deformability on the stability characteristics of the system, we consider here two separate cases : (i) $(Ga, \Sigma) = (0.1, 10^3)$, represents a liquid layer with a deformable free surface at the microgravity environment, and (ii) $(Ga, \Sigma) \to \infty$, refers to a liquid layer with non-deformable free surface.

## 3. Basic state and linear stability analysis

In this section, we present a linear stability analysis for small perturbations around the quiescent liquid film with laterally uniform temperature and concentration distribution. The purely conductive state of the system is represented by

$$v^o = \mathbf{0}, \qquad \tau^o = \mathbf{0}, \qquad h^o = 1, \qquad p^o = Ga(1-z),$$
$$T^o = 1 - z + Bi^{-1}, \qquad c^o = \chi z + const. \qquad (3.1a\text{-}f)$$

which are the steady-state solutions of $(2.8)-(2.10)$. One can notice that the elasticity of the fluid does not influence this basic state. We now study the stability of this basic state by introducing the following two-dimensional infinitesimal normal perturbations (denoted by a tilde) to the steady-state solutions (3.1),

$$v = v^o + \tilde{v}(x,z,t), \qquad \tau = \tau^o + \tilde{\tau}(x,z,t), \qquad p = p^o + \tilde{p}(x,z,t),$$
$$T = T^o + \tilde{\theta}(x,z,t), \qquad h = h^o + \tilde{\xi}(x,z,t), \qquad c = c^o + \tilde{c}(x,z,t). \qquad (3.2a\text{-}f)$$

Linearization of $(2.8)-(2.10)$ by neglecting the terms nonlinear in perturbations yields the following set of governing equations and boundary conditions:

$$\nabla \cdot \tilde{v} = 0, \qquad (3.3a)$$

$$Pr^{-1}\frac{\partial \tilde{\mathbf{v}}}{\partial t} = -\nabla \tilde{p} + \nabla \cdot \tilde{\tau}, \tag{3.3b}$$

$$\frac{\partial \tilde{\theta}}{\partial t} = \nabla^2 \tilde{\theta} + \tilde{w}, \tag{3.3c}$$

$$\frac{\partial \tilde{c}}{\partial t} + \chi \tilde{w} = Le\left(\nabla^2 \tilde{c} + \chi \nabla^2 \tilde{T}\right); \tag{3.3d}$$

$$\tilde{\mathbf{v}} = \mathbf{0}, \quad \frac{\partial \tilde{\theta}}{\partial z} = 0, \quad \frac{\partial \tilde{c}}{\partial z} = 0 \quad \text{at } z = 0, \tag{3.4a-c}$$

$$\frac{\partial \tilde{\xi}}{\partial t} = \tilde{w}, \quad \frac{\partial \tilde{\theta}}{\partial z} = -Bi\left(\tilde{\theta} - \tilde{\xi}\right), \quad \frac{\partial \tilde{c}}{\partial z} = \chi Bi\left(\tilde{\theta} - \tilde{\xi}\right),$$

$$\tilde{\tau}_{xz} = Ma \frac{\partial}{\partial x}\left(\tilde{c} - \tilde{\theta} + \tilde{\xi} + \chi \tilde{\xi}\right), \quad -\tilde{p} + Ga\, \tilde{\xi} + \tilde{\tau}_{zz} = \Sigma \frac{\partial^2 \tilde{\xi}}{\partial x^2} \quad \text{at } z = 1, \tag{3.5a-e}$$

whereas the constitutive equation (2.11) reads

$$\tilde{\tau} + De \frac{\partial \tilde{\tau}}{\partial t} = \left[(\nabla \tilde{\mathbf{v}}) + (\nabla \tilde{\mathbf{v}})^T\right]. \tag{3.6}$$

We now cast the BVP $(3.3)-(3.5)$ in terms of the stream function $\tilde{\psi}(x,z,t)$, so that

$$\tilde{u} = \frac{\partial \tilde{\psi}}{\partial z}, \quad \tilde{w} = -\frac{\partial \tilde{\psi}}{\partial x}. \tag{3.7a,b}$$

The basic idea behind the stream function formulation is to eliminate the pressure term $\tilde{p}$ from the system of equations $(3.3)-(3.5)$. Introducing relationships (3.7) and the constitutive equation for Maxwell viscoelastic model (3.6) into $(3.3)-(3.5)$, we finally arrive at:

$$Pr^{-1}\left(\frac{\partial}{\partial t}\nabla^2\tilde{\psi} + De\frac{\partial^2}{\partial t^2}\nabla^2\tilde{\psi}\right) = \nabla^4\tilde{\psi}, \tag{3.8a}$$

$$\frac{\partial \tilde{\theta}}{\partial t} = \nabla^2\tilde{\theta} - \frac{\partial \tilde{\psi}}{\partial x}, \tag{3.8b}$$

$$\frac{\partial \tilde{c}}{\partial t} - \chi \frac{\partial \tilde{\psi}}{\partial x} = Le\left[\nabla^2\tilde{c} + \chi \nabla^2\tilde{T}\right], \tag{3.8c}$$

with the boundary conditions,

$$\tilde{\psi} = 0, \quad \frac{\partial \tilde{\psi}}{\partial z} = 0, \quad \frac{\partial \tilde{\theta}}{\partial z} = 0, \quad \frac{\partial \tilde{c}}{\partial z} = 0 \quad \text{at } z = 0, \tag{3.9a-d}$$

$$\frac{\partial \tilde{\xi}}{\partial t} = -\frac{\partial \tilde{\psi}}{\partial x}, \quad \frac{\partial \tilde{\theta}}{\partial z} = -Bi\left(\tilde{\theta} - \tilde{\xi}\right), \quad \frac{\partial \tilde{c}}{\partial z} = \chi Bi\left(\tilde{\theta} - \tilde{\xi}\right), \tag{3.10a-c}$$

$$\frac{\partial^2 \tilde{\psi}}{\partial z^2} - \frac{\partial^2 \tilde{\psi}}{\partial x^2} = Ma\frac{\partial}{\partial x}\left(\tilde{c} - \tilde{\theta} + \tilde{\xi} + \chi \tilde{\xi}\right) + Ma\, De \frac{\partial^2}{\partial t \partial x}\left(\tilde{c} - \tilde{\theta} + \tilde{\xi} + \chi \tilde{\xi}\right), \tag{3.10d}$$

$$\left(1+De\frac{\partial}{\partial t}\right)\left(\Sigma\frac{\partial^3\tilde{\xi}}{\partial x^3}-Pr^{-1}\frac{\partial^2\tilde{\psi}}{\partial t\partial z}-Ga\frac{\partial\tilde{\xi}}{\partial x}\right)=-\frac{\partial}{\partial z}\left(3\frac{\partial^2}{\partial x^2}+\frac{\partial^2}{\partial z^2}\right)\tilde{\psi} \quad \text{at } z=1. \quad (3.10e)$$

Noticing that the basic state (3.1) is invariant with respect to $x$ and $t$, we use the Fourier decomposition to separate the $x$ and $t$ dependency of the perturbed fields ($\tilde{\psi}, \tilde{\theta}, \tilde{c}, \tilde{\xi}$) from that with $z$:

$$\left(\tilde{\psi}(x,z,t),\, \tilde{\theta}(x,z,t),\, \tilde{c}(x,z,t),\, \tilde{\xi}(x,z,t)\right)=\left(\hat{\psi}(z),\, \hat{\theta}(z),\, \hat{c}(z),\, \hat{\xi}(z)\right)\exp(i\,k\,x-\lambda\,t), \quad (3.11)$$

where ($\hat{\psi}, \hat{\theta}, \hat{c}, \hat{\xi}$) are the amplitudes of perturbations, $k$ denotes the dimensionless horizontal wavenumber and $\lambda=\Omega+i\omega$ refers to the decay rate of the perturbations. The parameter $\omega$ (a real quantity) represents the frequency of perturbation. Hence, the dynamics of these infinitesimal perturbations is now governed by the following eigenvalue problem (EVP):

$$Pr\frac{d^4\hat{\psi}}{dz^4}-\left(\lambda^2 De-\lambda+2Prk^2\right)\frac{d^2\hat{\psi}}{dz^2}+\left(\lambda^2 De-\lambda+Prk^2\right)k^2\hat{\psi}=0, \quad (3.12a)$$

$$\frac{d^2\hat{\theta}}{dz^2}+\left(\lambda-k^2\right)\hat{\theta}=i\,k\,\hat{\psi}, \quad (3.12b)$$

$$Le\frac{d^2\hat{c}}{dz^2}+\left(\lambda-Lek^2\right)\hat{c}=-Le\chi\left(\frac{d^2\hat{\theta}}{dz^2}-k^2\hat{\theta}\right)-i\chi k\hat{\psi}; \quad (3.12c)$$

$$\hat{\psi}=0,\quad \frac{d\hat{\psi}}{dz}=0,\quad \frac{d\hat{\theta}}{dz}=0,\quad \frac{d\hat{c}}{dz}=0 \quad \text{at } z=0, \quad (3.13a\text{-}d)$$

$$i\,k\,\hat{\psi}=\lambda\hat{\xi},\quad \frac{d\hat{\theta}}{dz}=-Bi\left(\hat{\theta}-\hat{\xi}\right),\quad \frac{d\hat{c}}{dz}=\chi Bi\left(\hat{\theta}-\hat{\xi}\right), \quad (3.14a\text{-}c)$$

$$\frac{d^2\hat{\psi}}{dz^2}+k^2\hat{\psi}=i\,Ma\,k\left(1-\lambda De\right)\left(\hat{c}-\hat{\theta}+\chi\hat{\xi}+\hat{\xi}\right), \quad (3.14d)$$

$$Pr\frac{d^3\hat{\psi}}{dz^3}+\left(\lambda-\lambda^2 De-3Prk^2\right)\frac{d\hat{\psi}}{dz}=i\,k\,Pr\left(1-\lambda De\right)\left(Ga+\Sigma k^2\right)\hat{\xi} \quad \text{at } z=1, \quad (3.14e)$$

with $\lambda$ and $Ma$ as the eigenvalues. This problem was recently analyzed by Sarma & Mondal (2019) for a *pure* viscoelastic film ($Le, \chi \to 0$). By surpassing this restriction, the more realistic binary aspect of the fluid is considered here, along with the incorporation of the Soret effect. Solving the system (3.12)−(3.14) for $\Omega=0$, now one can obtain the neutral stability curves, which demarcate the stable regime from the unstable one. However, the complexity of the solvability conditions here restrains us from taking an analytical approach. Therefore the EVP is solved numerically using the fourth-order Runge-Kutta method with shooting technique (Keller 2018) for disturbances with arbitrary values of $k$. An approximate model will be developed in § 6 in the asymptotic limit $k \to 0$.

It is well known that provided a good initial guess; the shooting method usually yields a more accurate solution at the reduced computational cost compared to most of the other numerical methods (see McFadden *et al.* 2010; Hirata *et al.* 2015). We have verified the accuracy of our numerical scheme by comparing the present results with those available in the literature and also with the results obtained from the approximate model. Confronted by the lack of published results on the thermosolutal Marangoni convection for viscoelastic fluids, we first test the accuracy of our numerical solution against the results of Shklyaev *et al.* (2009). These authors numerically investigated the instability phenomenon for a Newtonian binary mixture. In figure 2, one can see that an excellent quantitative agreement exists between the present results and the computations of Shklyaev *et al.* (2009) for both the values of *Bi*. The numerical results are also found to agree well with the results of the approximate model (see figures 14-16) for all but excluding the parameter values which violate the approximations considered to derive the model (discussed in § 6). These comparisons ensure the accuracy of the present numerical scheme for the entire parametric range of interest.

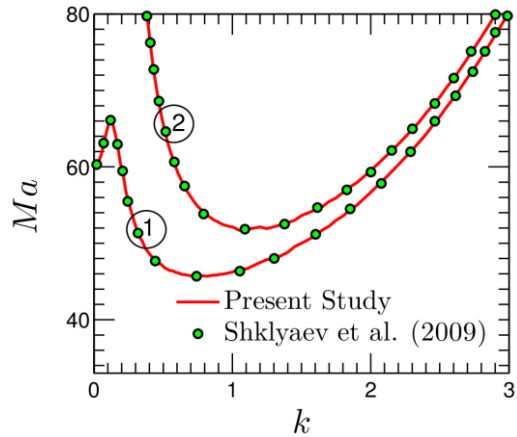

FIGURE 2. (Color online) Comparison of the present numerical result with the results of Shklyaev et al. (2009) (shown by marker "**o**") via the neutral stability curve at $Pr = 2$, $\chi = -0.2$, $Le = 10^{-3}$. Curves marked by 1 and 2 corresponds to $Bi = 0$ and $Bi = 0.1$ respectively. To represent the characteristics of a Newtonian binary liquid with a non-deformable free surface, we consider $De = 0$ and $(Ga, \Sigma) \to \infty$.

The EVP posed by (3.12)−(3.14) suggests the possible emergence of two different instability modes in the system: (i) monotonic mode (or stationary convection) and (ii) oscillatory mode (or overstability) for which disturbances grow with temporal oscillations. The stability thresholds for the monotonic and oscillatory modes can be obtained from (3.12)−(3.14) by substituting $\lambda = 0$ and $\lambda = i\omega$ respectively. Note that, to find the stability margin for the oscillatory instability mode, we numerically seek such value of $\omega$ for which the imaginary part of $Ma$ vanishes. Repetition of this procedure for a broad range of $k$ yields the neutral stability curve for this particular instability mode.

## 4. The linear stability picture

In this section, we analyze the stability picture obtained through numerical computations. Emphasis is put on understanding how viscoelasticity in the presence of Soret effect deviates the stability characteristics of the system from its Newtonian counterpart. For the convenience in analysis, we divide the entire disturbance spectrum into two different regimes: (i) long-wave regime, $k < O(1)$, and (ii) short-wave regime, $k \gtrsim O(1)$. Furthermore, it is important to remark that we fix $Pr = 10$ for all the graphical results. This is because the stability margin shows no substantial variation with $Pr$ against both the long-wave and short-wave perturbations (also apparent from the approximate model derived in § 6).

### 4.1. *Effect of elasticity and the free surface deformations*

Let us first start with the monotonic instability mode. Figure 3 plots the neutral stability curves for this particular instability mode. The solid line here represents the stability threshold for a liquid layer with a deformable free surface $(Ga, \Sigma) = (0.1, 10^3)$, while the dotted one depicts the stability boundary for a non-deformable surface $(Ga, \Sigma) \to \infty$. It can be observed that over the entire range of the disturbance wavenumber $k$, there exists a minimum value for the Marangoni number $Ma$ (indicated by the marker "o", see inset of figure 3) only above which the instability first sets in the system. We call this $Ma$ as the critical Marangoni number ($Ma_c$) and the corresponding $k$ and $\omega$ as the critical wavenumber ($k_c$) and critical oscillation frequency ($\omega_c$), respectively.

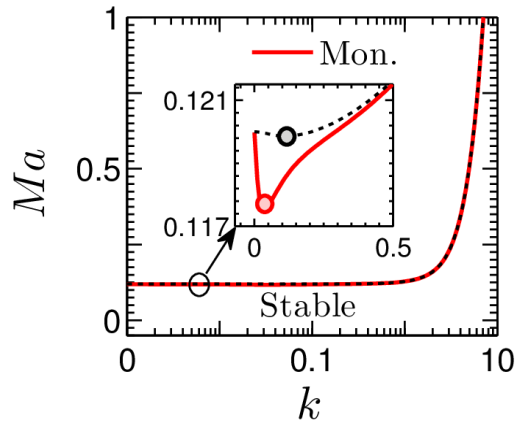

FIGURE 3. (Color online) Neutral stability curves $Ma$ ($k$) for the monotonic instability mode. The solid line represents the stability threshold for a deformable free surface $(Ga, \Sigma) = (0.1, 10^3)$, and the dotted line demonstrates the stability margin for a non-deformable free surface $(Ga, \Sigma) \to \infty$. The dot (o) mark on each neutral curve represents the critical point of the curve. Inset depicts the effect of the deformability of the free surface on the stability threshold in the long-wave regime. Other parameters: $Bi = 0.01$, $\chi = 0.5$, $Le = 10^{-3}$, $Pr = 10$.

From figure 3, it is clear that irrespective of the free surface deformability, the monotonic disturbances always emerge in the long-wave form (i.e. $k_c \ll O(1)$). Nevertheless, the increased gravitational and surface tension forces for a non-deformable surface slightly delay the onset of these long-wave disturbances in the system ($Ma_{c,(Ga,\Sigma)\to\infty} > Ma_{c,(Ga,\Sigma)\to(0.1,10^3)}$, see inset of figure 3). Notably, the elastic behaviour of the fluid does not influence the stability threshold for this instability mode. This is due to the vanishing of any temporal components for this stationary convection (cf. (3.12)−(3.14) substituting $\lambda = 0$). The role of other non-dimensional parameters on the stability margin for the monotonic instability mode will be discussed systematically in the subsequent subsections.

We now focus our attention on the disturbances that emerge with temporal oscillations ($\omega \neq 0$), giving rise to Hopf bifurcation. Previous investigations on the pure thermocapillary driven instability in a viscoelastic film (Lebon *et al.* 1994; Parmentier *et al.* 2000; Ramkissoon *et al.* 2006) suggest that oscillatory disturbances are more likely to appear in a highly viscoelastic film (highly and weakly viscoelastic fluids will be defined shortly). For the present thermo-solutal driven convection process, we will demonstrate that depending on the nature of the Soret coefficient (i.e. whether $\chi > 0$ or $\chi < 0$) two different oscillatory instabilities can emerge in the system. We call them the oscillatory-I and the oscillatory-II mode. It is important to remark that the characteristics of the oscillatory-I mode have been extensively studied in the literature in the context of a Newtonian binary mixture (Bestehorn & Borcia 2010; Podolny *et al.* 2005; Shklyaev *et al.* 2009; Skarda *et al.* 1998). However, its behaviour for a viscoelastic binary fluid has not been investigated yet. On the other hand, to our knowledge, *the oscillatory-II mode has entirely remained unexplored, even for a Newtonian binary fluid* (perhaps due to limited examination over the model parameters). We will demonstrate that for a viscoelastic binary mixture, while the oscillatory-I mode is more universal, the oscillatory-II instability can also get dominant in the system under appropriate model parameter values.

Figure 4 plots the neutral stability curves as well as the corresponding oscillation frequencies for the oscillatory-I mode. The solid and the dash-dotted lines represent the stability margin for a deformable free surface, and their adjacent dotted lines depict the stability boundary for a non-deformable free surface.

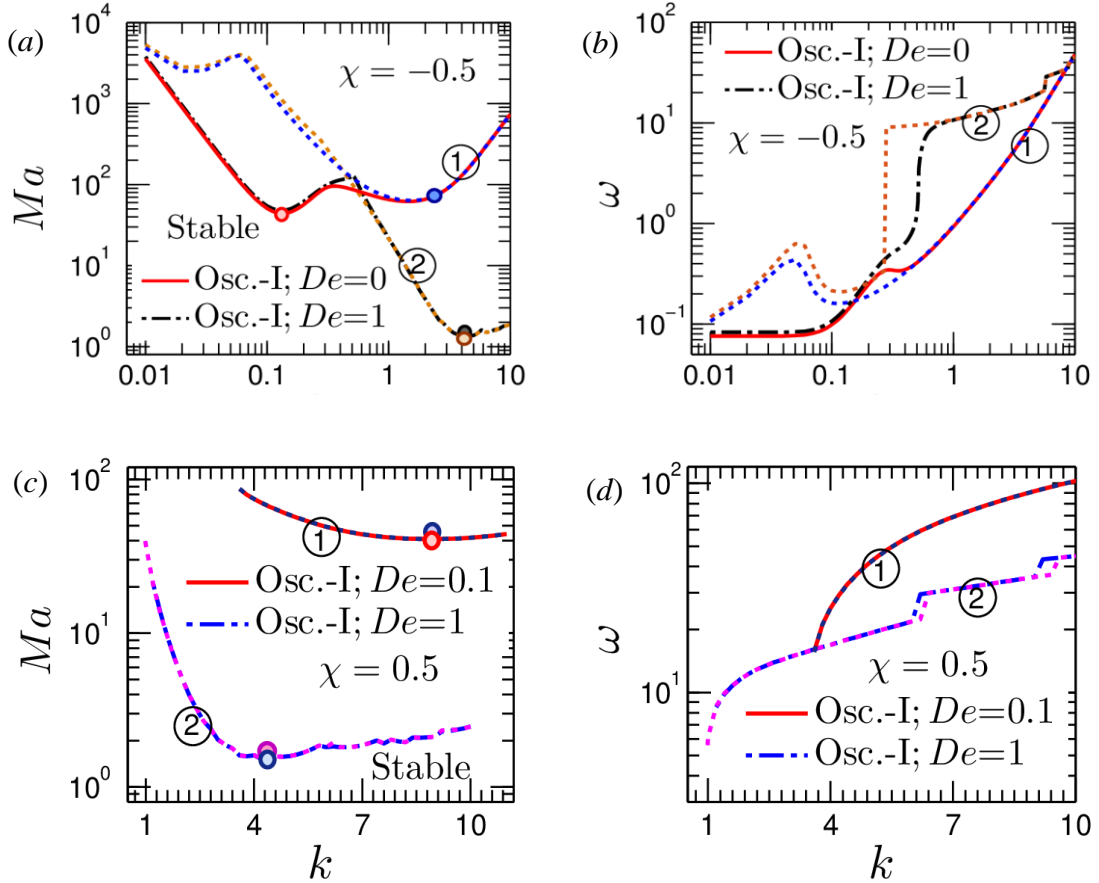

FIGURE 4. (Color online) (*a,c*) Neutral stability curves $Ma(k)$, and (*b,d*) the corresponding oscillation frequency $\omega$ for oscillatory-I instability mode for $\chi < 0$ $(=-0.5)$ and $\chi > 0$ $(=0.5)$ respectively. For (*a,b*): lines marked by 1 and 2 corresponds to $De=0$ and $De=1$ respectively; for (*c,d*): lines marked by 1 and 2 corresponds to $De=0.1$ and $De=1$ respectively. In each panel, the solid and the dash-dotted lines depict the results for a deformable free surface $(Ga,\Sigma) = (0.1,10^3)$; the adjacent dotted lines represents the results for a non-deformable free surface $(Ga,\Sigma) \to \infty$ for the same parameter values. The dot (o) mark on each neutral curve denotes the critical point (or the global minimum) of the curve. Other parameters: $Bi = 0.1$, $Le = 0.01$, $Pr = 10$.

Panel (*a*) demonstrates that for $\chi < 0$, the neutral curves consist of two branches, each characterized by a distinct local minimum. Note that of these two minima, while one resides in the long-wave regime $(k_c < O(1))$, the other lies in the short-wave regime $(k_c \gtrsim O(1))$.

Accordingly, we call these branches as the long-wave and the short-wave branch, respectively. It can be clearly seen that $Ma_c$ is a strong function of the deformability of free surface for the long-wave branch. Thus, similar to the monotonic mode, reduced deformability of the free surface enhances the stability of the system against the long-wave oscillatory-I disturbances as well. However, the onset of these disturbances essentially remains unaffected by the elastic behaviour of the fluid (see panel *a*, the long-wave branch for $De = 0$ and 1 merge into a single curve). On the other hand, $Ma_c$ for the short-wave branch is not influenced by the deformability of the free surface, rather governed by the elasticity of the mixture. Increased elasticity of the fluid here substantially promotes the onset of short-wave oscillatory-I instability in the system. This suggests that based on the deformability of the free and the elasticity level of the fluid, either of these branches can hold the position of global minimum (see the marker "o" on each neutral curve in panel *a*). In other words, the oscillatory-I disturbances can emerge both in the long-wave and short-wave modes for $\chi < 0$.

Panel (*c*) shows that the long-wave branch disappears for $\chi > 0$, leaving only the short-wave branch. In particular, in this regime of $\chi$, the oscillatory-I disturbances are found only for $De > 0$. This indicates that the emergence of short-wave oscillatory-I instability for $\chi > 0$ is the sole manifestation of the elastic behaviour of the fluid. Similarly to the case $\chi < 0$, the stability region remains unaltered by the deformability of the free surface but diminishes drastically with the increasing elasticity of the mixture.

The oscillation frequency $\omega$ of the neutral perturbations corresponding to each neutral curve of panels (*a,c*) is plotted in panels (*b,d*). The deformability of the free surface controls $\omega$ only in the long-wave regime, while, $\omega$ is primarily modulated by the elasticity of the fluid in the short-wave regime. It should be further noted that although the $\omega(k)$ variation is smooth in the long-wave regime, discontinuity appears in the short-wave regime (particularly at higher $De$ values). This behaviour is also reflected in the neutral curves at large $De$, wherein the curves become slightly non-monotonic in the short-wave regime, particularly at higher values of $k$ (see panels *a,c*). Similar features of the neutral curves have been previously reported by Dauby *et al.* (1993) and Parmentier *et al.* (2000) in the context of pure thermocapillary driven convection in a viscoelastic film.

It is now clear that for $\chi < 0$, the oscillatory-I disturbances can develop both in the Newtonian and viscoelastic binary mixtures. However, the elasticity of the fluid significantly influences the onset of this particular instability mode in the system. To illustrate this elasticity-based transition of the stability picture, we plot in figure 5 the critical Marangoni number, $Ma_c$, the corresponding wavenumber, $k_c$, and oscillation frequency, $\omega_c$ as functions of Deborah number, $De$ for both the cases of deformable (solid line) and non-deformable (dotted line) free surface. Note that, $Ma_c$ refers here to the global minimum of the oscillatory-I neutral curve. Two regimes are clearly distinguishable from the variations depicted by figure 5: a weakly elastic regime (for $De \lesssim 0.1$) wherein the stability behaviour resembles that of a Newtonian binary fluid (at least for bifurcation around the conductive base state), and a strong elastic regime (for $De > 0.1$) where the elasticity of the fluid governs the stability threshold and the critical parameters $(k_c, \omega_c)$. The transition between these two regimes is marked by sharp discontinuities in $k_c$ and $\omega_c$ (see the arrow mark in panels $b,c$).

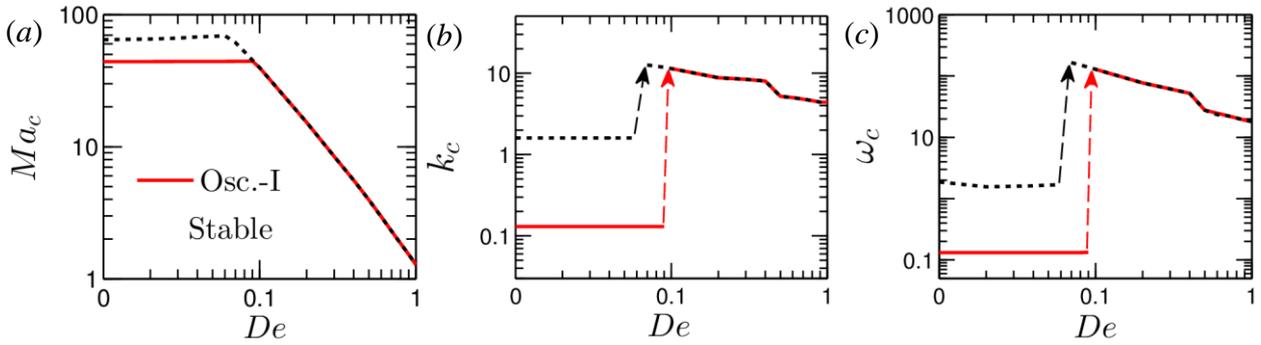

FIGURE 5. (Color online) Variation of the ($a$) critical Marangoni number $Ma_c$, and the corresponding critical ($b$) wavenumber $k_c$, and the ($c$) oscillation frequency $\omega_c$ with Deborah number $De$ for the oscillatory-I instability mode for $\chi < 0$. The solid line depicts the variation for a deformable free surface $(Ga, \Sigma) = (0.1, 10^3)$; and the dotted one for a non-deformable surface $(Ga, \Sigma) \to \infty$. The arrow marks in panels ($b,c$) illustrate a switchover in the instability behaviour with the increasing elasticity of the fluid. Other parameters: $Bi = 0.1$, $Le = 0.01$, $\chi = -0.5$.

A key observation from figure 5 is that, while the critical parameters $(Ma_c, k_c, \omega_c)$ are governed by the deformability of the free surface rather than the elasticity of the fluid in the weakly elastic regime, the opposite is true for the highly elastic regime. Panels ($a$) shows that the reducing deformability of the free surface dampens the onset of oscillatory-I instability for $De \lesssim 0.1$. The resulting disturbances emerge in the long-wave form ($k_c \approx 0.1$) for a deformable free surface and the short-wave form ($k_c \approx 1$) for a non-deformable free surface (see panel $b$).

On the other hand, for a highly viscoelastic mixture ($De > 0.1$), irrespective of the free surface deformability, the disturbances always set in the short-wavelength form. An inverse variation of the parameters ($Ma_c, k_c, \omega_c$) with $De$ in this regime suggests that, with enhanced elasticity of the binary mixture, the conductive state is more likely to bifurcate to the short-wave oscillatory-I mode with the more easily detectable convective pattern.

Another interesting feature presented by figure 5 is that, for $De \approx 0.1$ (the boundary separating the weakly and highly elastic regime for a liquid layer with deformable free surface), $Ma_c$ for the onset of long-wave and short-wave oscillatory-I perturbations coincide. Therefore, a competition between the respective instability modes can take place in the system for $(De, Ga, \Sigma) \approx (0.1, 0.1, 10^3)$.

### 4.2. *Role of the thermocapillary and the solutocapillary effects*

In this subsection, we investigate the contributions of the thermocapillary and solutocapillary forces on the development of instabilities in the system. This is done by plotting the neutral stability curves for both the monotonic and oscillatory modes for different values of $\chi$ in figure 6. Recall $\chi = 0$ refers in this analysis to the case of pure thermocapillary driven convection. Figure 6(*a*) shows that the domain of stability reduces substantially as $\chi$ increases from zero. This suggests that both thermocapillary and solutocapillary forces play a destabilizing role in the emergence of monotonic instability for $\chi > 0$. However, it is important to note that $k_c$ for this instability mode is not decided by the solutal effects (at least for a deformable surface, the case of the non-deformable surface will be discussed in figure 7). Although not shown here graphically, this particular instability mode can appear in the system even for $\chi < 0$, within a narrow interval of $\chi$. This range will be identified in § 6, performing a long-wave asymptotic analysis of the problem.

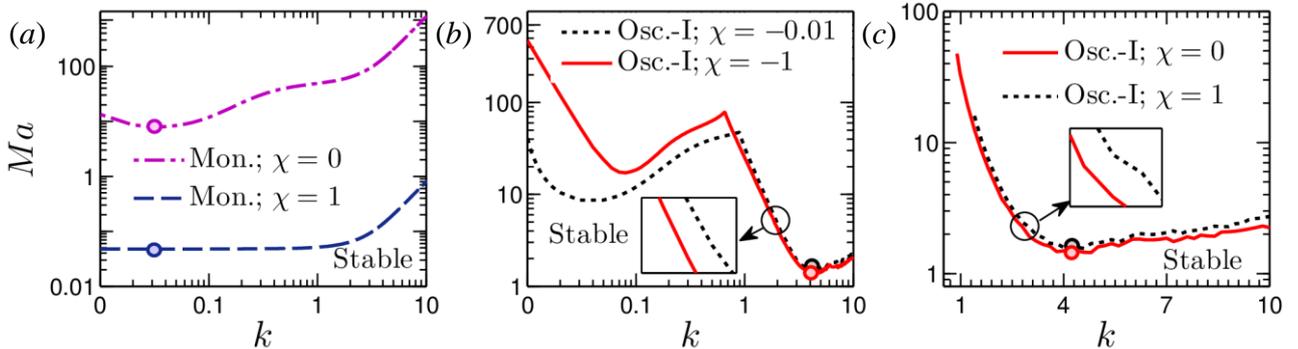

FIGURE 6. (Color online) Neutral stability curves $Ma(k)$ for the monotonic and oscillatory–I modes in the (*a,c*) positive, and (*b*) negative Soret number $\chi$ domains. The long-wave branch for the oscillatory-I mode emerges only when $\chi < 0$. The dot (o) mark on each neutral curve represents the critical point of the curve. Other parameters: $Bi = 0.01$, $De = 1$, $Le = 10^{-3}$, $Ga = 0.1$, $\Sigma = 10^3$.

Figure 6(*b*) demonstrates that for the long-wave branch of the oscillatory-I mode, the solutocapillary acts stabilizing, while the thermocapillary turns into the destabilizing mechanism. Such opposite contributions of the driving forces give rise to the long-wave oscillatory-I perturbations. On the other hand, thermocapillarity, coupled with the elasticity of the fluid, primarily give rise to the short-wave disturbances. Solutocapillarity provides here only a small correction to the stability margin. Interestingly, for this short-wave branch, an increasing $|\chi|$ in the range $\chi < 0$ weakly destabilizes the system, whereas the increment in $\chi$ for $\chi > 0$ leads to a mild stabilization of the system (see inset, figures 6*b,c*).

Figure 7 plots the variations of the critical Marangoni number $Ma_c$ and the corresponding wavenumber $k_c$ with $\chi$ for both the cases of the deformable and non-deformable free surface. Clearly, an increase in $\chi$ leads to a strong destabilization of the system with respect to the monotonic disturbances (see figure 7*a*). It should be noted that, for this instability mode, although the deformability of free surface weakly influences the stability threshold (see figure 3, this is not even perceptible in the scale of figure 7*a*); nevertheless, it plays an important role in determining the size of the convection cells. In figure 7(b) one can see that $k_{c,(Ga,\Sigma)\to\infty} > k_{c,(Ga,\Sigma)=(0.1,10^3)}$, implying a highly deformable free surface allows the formation of much larger sized stationary convective patterns compared to its non-deformable counterpart.

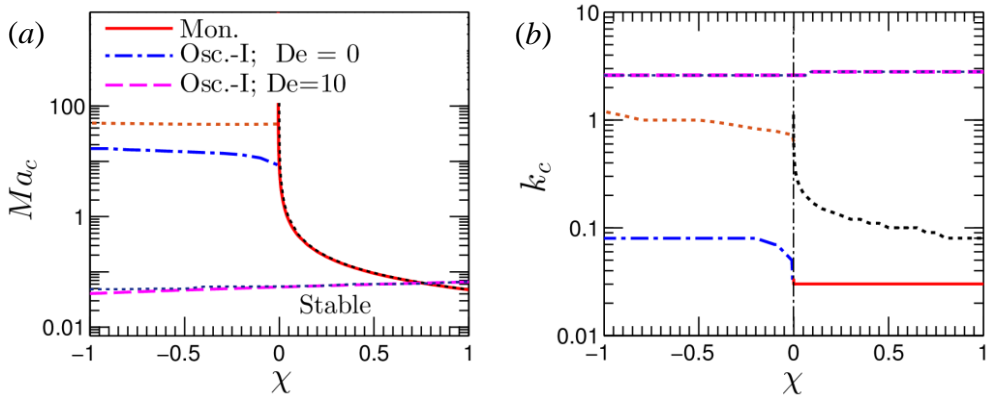

FIGURE 7. (Color online) Variation of the (*a*) critical Marangoni number $Ma_c$, and the corresponding (*b*) critical wavenumber $k_c$ with $\chi$. In each panel, the solid, dash-dotted, and dashed lines represent the variations for a deformable free surface $(Ga, \Sigma) = (0.1, 10^3)$. The dotted lines adjacent to each instability mode represent the same variations for a non-deformable free surface $(Ga, \Sigma) \to \infty$. Other parameters: $Bi = 0.01$, $Le = 10^{-3}$.

For the oscillatory-I mode, figure 7(a) shows that an increasing $|\chi|$ augments the stability region for a Newtonian binary liquid film with a deformable free surface. However, for a non-deformable free surface (or for a highly viscoelastic mixture irrespective of the free surface deformability), the stability thresholds found to remain nearly independent of $\chi$. In this regard, it is important to observe in figure 7(b) that, for a Newtonian binary mixture with a deformable surface $k_c$ lies in the long-wave regime, whereas, for a non-deformable surface (or for a highly viscoelastic mixture) $k_c$ resides in the short-wave regime. The fact that the solutocapillary effect is dominant only in the long-wave regime (see figure 6) explains the variations in figure 7(a).

Before concluding this subsection, an additional remark about figure 7(a) is necessary. Note that, since $Ma_{c,mon} = Ma_{c,osc.-I}$ at the intersection point between the neutral curves for the monotonic and oscillatory-I modes, a competition between the respective instability modes can, therefore, occur for $\chi$ values corresponding to this point. Now, as the oscillatory-I disturbances can appear for any value of $\chi$ for $De > 0$, and their onset gets triggered by the increasing elasticity of the fluid, the neutral curve for the monotonic mode presented in figure 7(a) is essentially the locus of such codimension-two bifurcation points. Towards the left of this curve, the conductive state bifurcates into the oscillatory-I mode, while a steady bifurcation (i.e. monotonic instability) takes place on to its right.

### 4.3. *Role of the thermal and solutal diffusivities*

Let us now discuss the influence of thermal and solute diffusivities on the onset of instability in the system. Here, we will demonstrate that based on the diffusivity ratio $Le(=D/\alpha)$, a different kind of oscillatory instability (i.e. oscillatory-II) can emerge in the fluid layer. Before that, we first investigate the role played by *Le* on the emergence of oscillatory-I disturbances.

Figure 8(a) shows that an increased solute diffusivity corresponding to higher values of *Le* enhances the stability of the system against the long-wave oscillatory-I disturbances. This is due to the stabilizing role of solutocapillarity (see figure 6*b*) in producing this instability mode. Similarly, a destabilizing (stabilizing) solutocapillary force for $\chi < 0$ $(\chi > 0)$ in the short-wave regime leads to a reduction (enhancement) in the stability threshold for the higher solute diffusivity (cf. figures 8*a,b* with figures 6*b,c*).

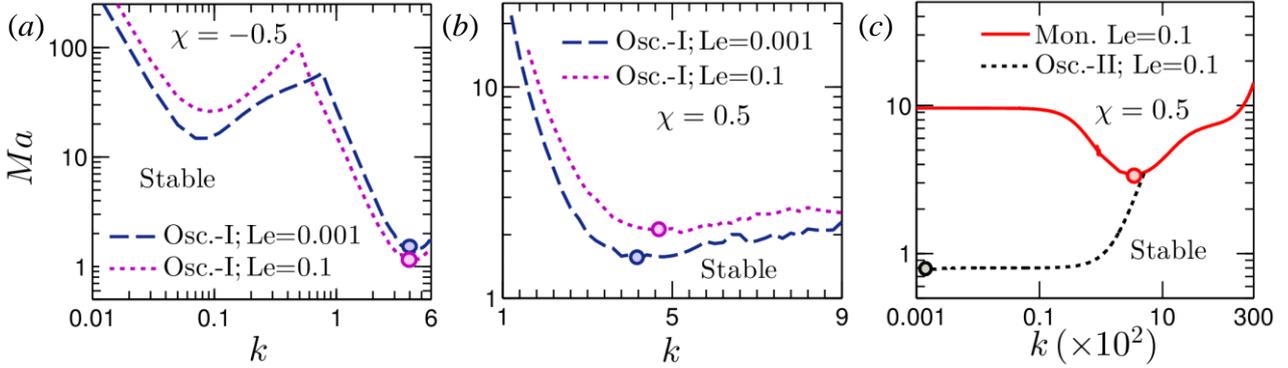

FIGURE 8. (Color online) Effect of Lewis number *Le* on the monotonic and oscillatory instability threshold for (*a,c*) $\chi < 0 (= -0.5)$, and (*b*) $\chi > 0 (= 0.5)$. The dot (○) mark on each neutral curve represents the critical point of the curve. Panel (*c*) shows that for a deformable free surface at higher *Le*, a different type of oscillatory instability (oscillatory-II) can emerge in the long-wave regime that merges with the neutral curve for monotonic mode at higher values of *k*. Other parameters: $Bi = 0.01$, $De = 1$, $Ga = 0.1$ and $\Sigma = 10^3$.

Figure 8(*c*) shows that for a sufficiently large *Le* ($\gtrsim O(10^{-2})$), the oscillatory-II disturbances can emerge in systems with a deformable free surface. Note that, at higher values *k*, the neutral curve for this particular instability mode (the dotted line) merges with the neutral curve for the monotonic mode (the solid line). This limits the appearance of oscillatory-II disturbances only in the long-wave form.

The other features of this instability mode are also found to be quite different from the oscillatory-I mode. First, while the oscillatory-I instability can emerge in the system for the entire permissible range of the model parameters, the oscillatory-II instability appears only in the case of a deformable free surface, for $\chi > 0$ and $Le \gtrsim O(10^{-2})$. Although not shown here graphically, such disturbances get damped with the reducing deformability of the free surface and eventually disappear in a non-deformable free surface. Second, the oscillation frequency of the oscillatory-II mode is several orders of magnitude smaller than the oscillatory-I mode (see figure 9). This suggests that the oscillatory-II perturbations develop with a significantly large oscillation period compared to the oscillatory-I disturbances.

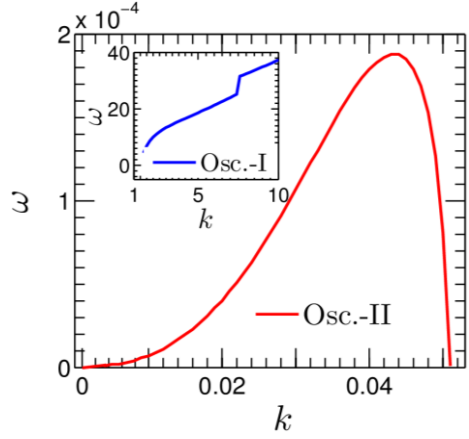

FIGURE 9. (Color online) Comparison of the oscillation frequency of neutral perturbations for oscillatory–II mode with oscillatory–I mode (shown in the inset). $\chi = 0.5$, $Bi = 0.01$, $De = 1$, $Le = 0.1$, $Ga = 0.1$, $\Sigma = 10^3$.

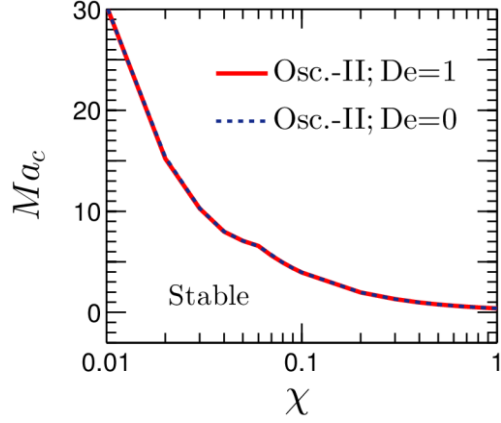

FIGURE 10. (Color online) Variation of the critical Marangoni number $Ma_c$ with $\chi$ for the oscillatory–II instability mode at $Bi = 0.01$, $Le = 0.1$ $Ga = 0.1$, $\Sigma = 10^3$. Note that no oscillatory–II instability emerges for $\chi \leq 0$.

Now, in order to understand the physical mechanism behind the origination of the oscillatory-II disturbances and elucidate the role of elasticity on their onset, we plot in figure 10 the variation of $Ma_c(\chi)$ with $De$ for this particular instability mode. The disappearance of these disturbances for purely thermocapillary driven convection ($Ma_c \to \infty$ as $\chi \to 0$) and the reduction of $Ma_c$ with $\chi$ suggests that the competition between the stabilizing thermocapillary and the destabilizing solutocapillary forces give rise to the oscillatory-II mode. A shorter mass diffusion time scale $(H^2/D)$ is therefore essential here to overcome the stabilizing action of thermocapillary by the destabilizing solutocapillary force. This explains the reason behind the emergence of oscillatory-II instability only for higher values of $Le\left(=\dfrac{H^2}{\alpha}\bigg/\dfrac{H^2}{D} \gtrsim O(10^{-2})\right)$.

Another key observation from figure 10 is that the $Ma_c(\chi)$ neutral curves for different $De$ values collapse into a single curve. This implies that the stability threshold for the oscillatory-II mode is not affected by the elastic behaviour of the binary mixture.

### 4.4. *Role of the heat transfer rate at the free surface*

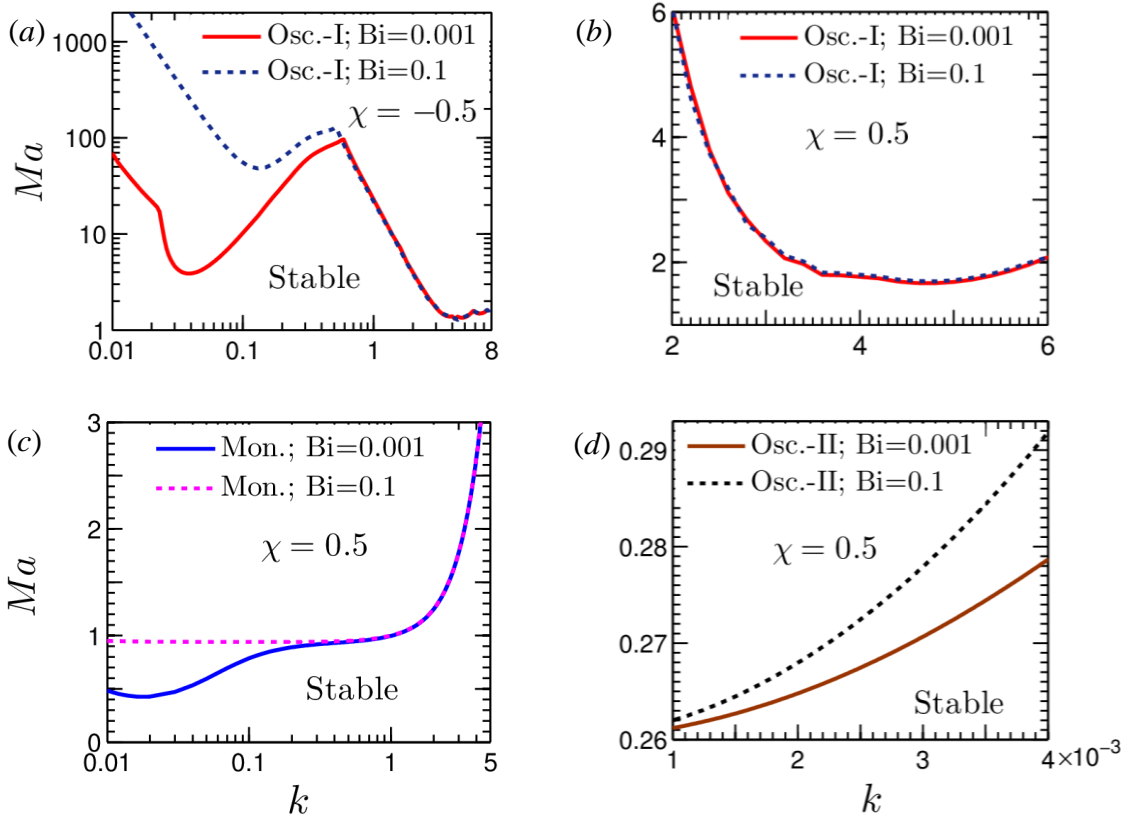

FIGURE 11. (Color online) Effect of Biot number $Bi$ on the monotonic and oscillatory instability threshold for (a) $\chi < 0 (= -0.5)$, and (b,c,d) $\chi > 0 (= 0.5)$ for a deformable free surface $(Ga, \Sigma) = (0.1, 10^3)$. Other parameters: $De = 1$, $Le = 0.01$.

Lastly, we discuss the role of the Biot number on the stability threshold of the system. In figure 11, the neutral stability curves for each instability mode are plotted for two different values of $Bi$ ($=10^{-3}$ and 0.1). It turns out that at higher values of $Bi$, the enhanced heat transfer rate from the free surface increases the stability of the system against the long-wave disturbances (see panels (a,c,d) at small $k$). However, the influence of $Bi$ is less significant in the short-wave regime, as indicated by the saturation of the curves in this regime (see panels (a-c) at large $k$). The magnitude of $Bi$, therefore, bears significant importance in the emergence of long-wave instability in the fluid layer.

## 5. Spatial structure of eigenvectors at neutral stability

We now briefly examine the effect of elasticity on the spatial structure of the eigenvectors $\hat{\psi}$, $\hat{\theta}$ and $\hat{c}$ at the neutral stability boundary. Figure 12 plots the normalized spatial profiles of $\hat{\psi}$ for each instability mode at two different levels of elasticity ($De = 0$ and 1) of the binary mixture.

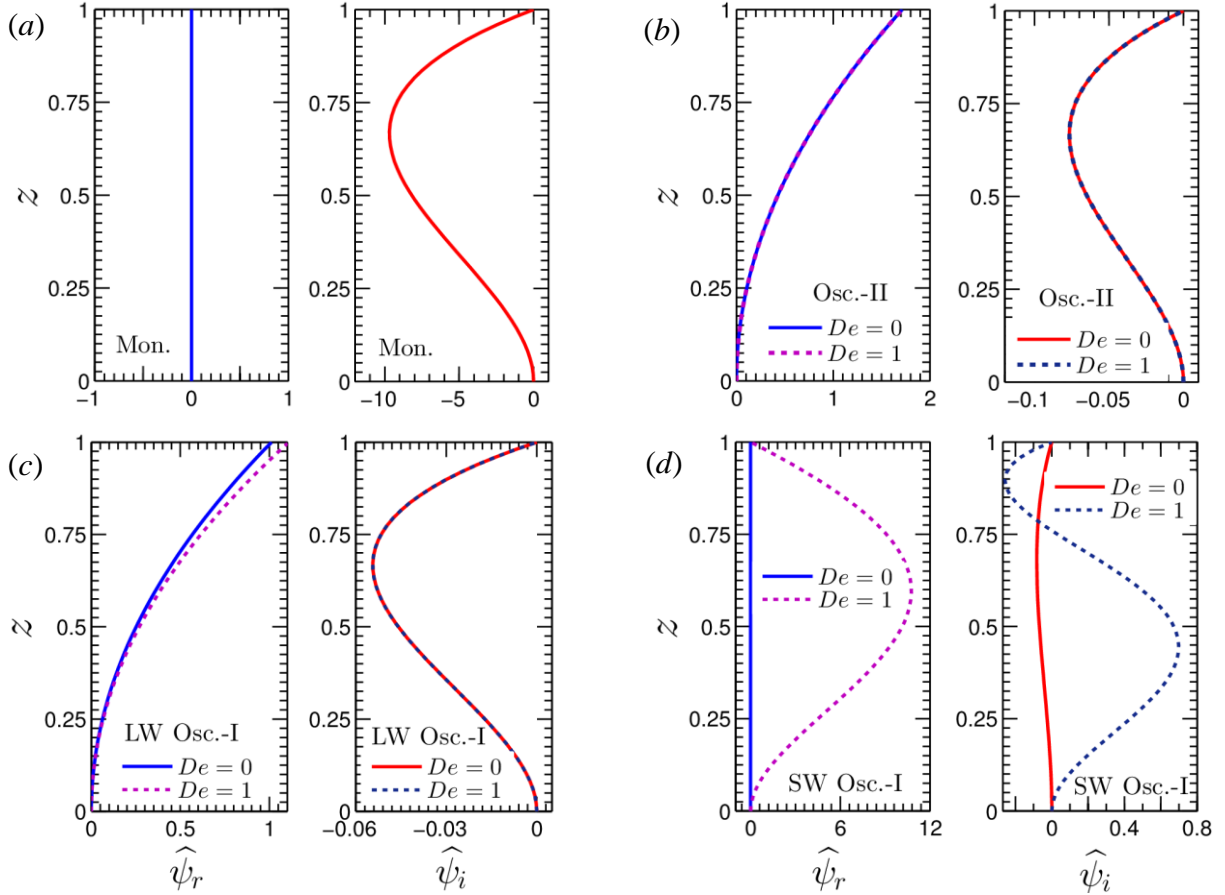

FIGURE 12. (Color online) Effect of elasticity on the eigenvector $\hat{\psi}$ profiles (normalized) at the neutral stability for: (*a*) monotonic mode, (*b*) oscillatory-II mode, (*c*) long-wave oscillatory –I mode, and (*d*) short-wave oscillatory –I mode. Profiles in panels (*a,b*) correspond to the critical point of the respective neutral curve presented in figure 8(*a*). Profiles in panels (*c,d*) refer to the critical point of the long-wave and short-wave branch of the neutral curves depicted in figure 4(*a*). The eigenvectors are plotted for a deformable free surface $(Ga, \Sigma) = (0.1, 10^3)$.

Note that, $\hat{\psi}(=\hat{\psi}_r + i\hat{\psi}_i)$ is a complex-valued eigenvector. For the monotonic instability mode, $\hat{\psi}_r = 0$ and thus, $\hat{\psi}(=i\hat{\psi}_i)$ assumes a purely imaginary value (see panel *a*). Similarly to the monotonic instability threshold, the spatial structure of $\hat{\psi}$ also remains independent of the elasticity of the mixture. For the oscillatory-II as well as the long-wave oscillatory-I mode, panels (*b,c*) demonstrate that the spatial structure of $\hat{\psi}$ are identical and essentially remains unaltered by the elasticity of the fluid. This observation is also consistent with the results obtained in § 4 (which suggest that the onset of long-wave oscillatory disturbances get least affected by the elastic behaviour of the binary mixture). On the other hand, the spatial shape of $\hat{\psi}$ for the short-wave oscillatory-I mode is found to be very sensitive to the elasticity of the fluid (see panel *d*). $\hat{\psi}$ exhibits here substantial spatial distortions with sharp gradients for higher values of *De*, yielding more complicated structures.

The spatial structures of the eigenvectors $\hat{\theta}(=\hat{\theta}_r + i\hat{\theta}_i)$ and $\hat{c}(=\hat{c}_r + i\hat{c}_i)$ for the short-wave oscillatory-I mode are presented in figure 13. Clearly, both the eigenvectors demonstrate more distorted spatial structures for increasing the level of elasticity of the mixture.

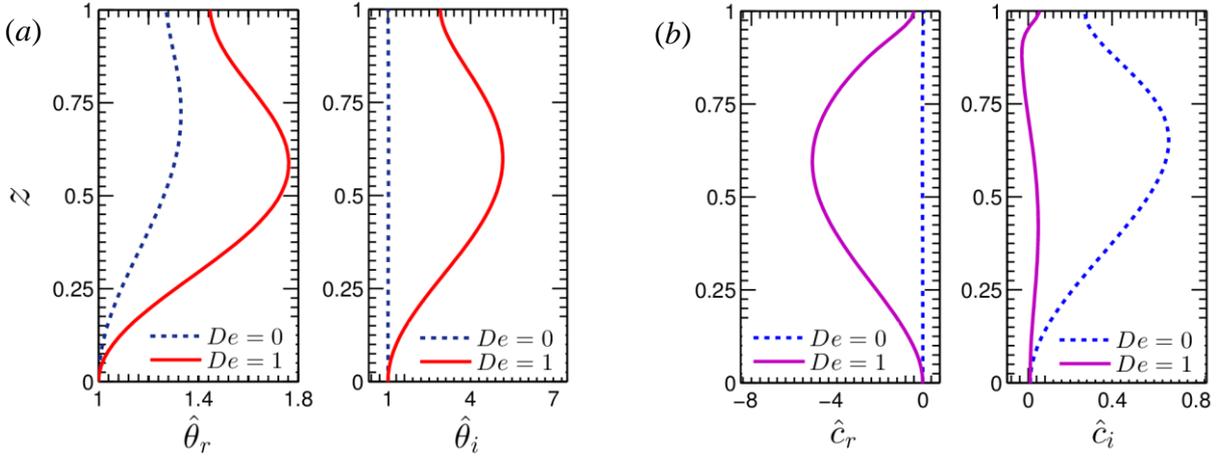

FIGURE 13. (Color online) Effect of elasticity on the spatial profiles of the (*a*) temperature $\hat{\theta}$ and (*b*) concentration $\hat{c}$ eigenvectors (normalized) at neutral stability. Profiles for each *De* in panels (*a,b*) correspond to the critical point of the short-wave oscillatory-I branch presented in figure 4(*a*).

We can, therefore, conclude that except for the short-wave oscillatory-I mode, the spatial structure of the eigenvectors for the remaining instability modes is not influenced by the elastic behaviour of the fluid.

## 6. An approximate model

In this section, we derive an approximate model performing long-wave asymptotic expansion of the EVP (3.12)−(3.14) and rescaling the parameter space $(Bi, De, \Sigma)$. This model can help to get a qualitative insight into the stability picture without numerically solving the eigenvalue problem.

In the long-wavelength limit, given the very small ratio between the mean film thickness $H$ and the disturbance wavelength $\ell$ (i.e. $\varepsilon = H/\ell \ll 1$), the horizontal variations evolve much slowly compared to the vertical ones. Here, we can apply the lubrication approximation, consisting of slow longitudinal $X \sim \varepsilon x$ and temporal $\mathcal{T} \sim \varepsilon^2 t$ variables. We, therefore, proceed with introducing the following scaling for $k$ and $\lambda$:

$$k = \varepsilon q, \quad \lambda = \varepsilon^2 \lambda_o. \qquad (6.1a,b)$$

Furthermore, for this analysis, we rescale the parameters $Bi$, $De$ and $\Sigma$ as

$$Bi = \varepsilon^2 \mathcal{B}, \quad De = \varepsilon^{-2} \mathcal{D}e, \quad \Sigma = \varepsilon^{-2} Ca. \qquad (6.2a\text{-}c)$$

Scaling (6.2) suggests that the proposed model remains effective only for the small values of $Bi$ and large $De$ and $\Sigma$. Recall, a small $Bi$ physically represents a poorly conducting free surface, while large $De$ signifies a highly viscoelastic fluid. It is important to note that owing to the weak influence of elasticity on the long-wave instability threshold (see figures 4,10), the projected model would be able to predict the stability margin in the long-wave regime even for small values of $De$ (graphically demonstrated a little later). The scaling (6.2b) will only help in better predicting the stability boundary in the short-wave regime for higher values of $De$ (surpassing the approximations made in (6.1)). Furthermore, the consideration of large $\Sigma$ is well justified here since its magnitude is usually high. However, we do not impose any restrictions on the magnitude of other dimensionless parameters, and they remain at $O(1)$ with respect to $\varepsilon$.

The perturbation fields are now expanded for $\varepsilon$ as follows:

$$\widehat{\psi} = \varepsilon \psi_o + \varepsilon^3 \psi_2 + ..., \quad \widehat{\theta} = T_o + \varepsilon^2 T_2 + ..., \quad \widehat{c} = c_o + \varepsilon^2 c_2 + ..., \quad \widehat{\xi} = \xi_o + \varepsilon^2 \xi_2 + .... \quad (6.3a\text{-}d)$$

Introducing the rescaled parameters and the expansions $(6.1)-(6.3)$ into the EVP $(3.12)$ – $(3.14)$, we collect the terms with identical order in $\varepsilon$. At the leading order, the governing equations (3.12a-c) simplify to

$$\frac{d^4 \psi_o}{dz^4} = 0, \quad \frac{d^2 T_o}{dz^2} = 0, \quad \frac{d^2 c_o}{dz^2} = 0; \qquad (6.4a\text{-}c)$$

and accompanied by the following set of boundary conditions:

$$\psi_o = 0, \quad \frac{d\psi_o}{dz} = 0, \quad \frac{dT_o}{dz} = 0, \quad \frac{dc_o}{dz} = 0 \quad \text{at } z = 0, \qquad (6.5a\text{-}d)$$

$$iq\psi_o = \lambda_o \xi_o, \quad \frac{dT_o}{dz} = 0, \quad \frac{dc_o}{dz} = 0, \qquad (6.6a\text{-}c)$$

$$\frac{d^3\psi_o}{dz^3} = iq\xi_o\left(Ga + Ca\,q^2\right)\left(1 - \lambda_o De\right), \qquad (6.6d)$$

$$\frac{d^2\psi_o}{dz^2} = iqMa\left(1 - \lambda_o De\right)\left(c_o - T_o + \xi_o + \chi\xi_o\right) \quad \text{at } z = 1. \qquad (6.6e)$$

The solutions to this BVP $(6.4)-(6.6)$ is given by

$$\psi_o = \frac{i\xi_o}{6q}\left[q^2\left(Ga + Ca\,q^2\right)\left(1 - \lambda_o De\right)\left(z^3 - z^2\right) - 6\lambda_o z^2\right], \quad T_o = \mathcal{J}, \quad c_o = \varphi, \quad (6.7a\text{-}c)$$

where $\mathcal{J}$ and $\varphi$ are constants yet to be determined.

At the second order with respect to $\varepsilon$, only the energy and mass balance equations find relevance. They read

$$\frac{d^2 T_2}{dz^2} + \left(\lambda_o - q^2\right)T_o = iq\psi_o, \qquad (6.8a)$$

$$Le\frac{d^2 c_2}{dz^2} + \left(\lambda_o - Le\,q^2\right)c_o = -i\chi(1 + Le)q\psi_o + Le\chi\lambda_o T_o. \qquad (6.8b)$$

The associated boundary conditions $(3.13c,d)$ and $(3.14b,c)$ now take the following form

$$\frac{dT_2}{dz} = 0, \quad \frac{dc_2}{dz} = 0 \quad \text{at,} \quad z = 0 \qquad (6.9a,b)$$

$$\frac{dT_2}{dz} = -\mathcal{B}(T_o - \xi_o), \quad \frac{dc_2}{dz} = \chi\mathcal{B}(T_o - \xi_o) \quad \text{at } z = 1. \qquad (6.10a,b)$$

Integrating $(6.8a,b)$ across the film $0 \leq z \leq 1$ and incorporating the boundary conditions $(6.9)$-$(6.10)$, we obtain

$$\xi_o = \frac{72\left(\phi + q^2\right)}{72\mathcal{B} - 24\lambda_o - \mathcal{G}\mathcal{Q}q^2}\mathcal{J}, \qquad (6.11)$$

$$\varphi = \frac{\chi\phi\left(72\mathcal{B} - 24\lambda_o - \mathcal{G}\mathcal{Q}q^2\right) + \left[\chi\left(1 + Le^{-1}\right)\left(24\lambda_o + \mathcal{G}\mathcal{Q}q^2\right) - 72\mathcal{B}\chi\right]\left(\phi + q^2\right)}{\left(72\mathcal{B} - 24\lambda_o - \mathcal{G}\mathcal{Q}q^2\right)\left(q^2 - \lambda_o Le^{-1}\right)}\mathcal{J}, \quad (6.12)$$

where $\phi = \mathcal{B} - \lambda_o$, $\mathcal{G} = \left(1 - \lambda_o De\right)$ and $\mathcal{Q} = Ga + Ca\,q^2$.

Finally, the substitution of $\xi_o$ and $\varphi$ into the tangential stress balance boundary condition $(6.6e)$ yields the following sought expression for $Ma$:

$$Ma = \frac{24}{q^2}\frac{\mathcal{N}_0 + \lambda_o\mathcal{N}_1 + \lambda_o^2\mathcal{N}_2 + \lambda_o^3\mathcal{N}_3}{\mathcal{D}_0 + \lambda_o\mathcal{D}_1 + \lambda_o^2\mathcal{D}_2 + \lambda_o^3\mathcal{D}_3}, \qquad (6.13)$$

the coefficients $\mathcal{N}_j$ and $\mathcal{D}_j$ $(j = 0-3)$ are defined in appendix A.

Equation (6.13) governs the stability threshold of the system for both the monotonic and the oscillatory disturbances within the approximations mentioned in (6.1)–(6.2). The validity bound for this analysis will be discussed in the forthcoming subsections. An inspection of (6.13) reveals that, in accordance with the numerical results presented in § 4, the stability threshold for both the instability modes are independent of *Pr*.

### 6.1. *Monotonic mode*

Let us start with the case of monotonic instability. Substitution of $\lambda_o = 0$ in (6.13) yields the explicit expression for the neutral stability curve for this instability mode:

$$Ma_{\text{mon.}} = \frac{48 Le \, \mathcal{Q}(\mathcal{B}+q^2)}{\mathcal{B}\chi\mathcal{Q}+\left[\mathcal{Q}\chi + Le(72+\mathcal{Q})(1+\chi)\right]q^2}. \tag{6.14}$$

Returning to the unscaled parameters $k$, $Bi$ and $\Sigma$ we get

$$Ma_{\text{mon.}} = \frac{48 Le\left(Ga+\Sigma k^2\right)\left(Bi+k^2\right)}{Bi\chi\left(Ga+\Sigma k^2\right)+\left[\chi\left(Ga+\Sigma k^2\right)+Le(1+\chi)\left(72+Ga+\Sigma k^2\right)\right]k^2}. \tag{6.15}$$

Equation (6.15) depicts the stability boundary for a system with a deformable free surface. For a non-deformable surface (i.e. in the limit $(Ga+\Sigma k^2)\to\infty$), the stability margin (6.15) becomes

$$Ma_{\text{mon.}} = \frac{48 Le\left(Bi+k^2\right)}{Bi\chi+(Le+\chi+Le\chi)k^2}. \tag{6.16}$$

The minimization of (6.15) for *Ma* now yields the following expression for the critical wavenumber $k_c$

$$k_c = \sqrt{\frac{Bi\,Ga\,\Sigma + \sqrt{72\,Bi\,Ga\,\Sigma(72+Ga-Bi\Sigma)}}{\Sigma(72-Bi\Sigma)}} \quad \text{for } \chi \neq -1. \tag{6.17}$$

Interestingly, equation (6.17) also indicates the validity domain of this approximate model. Note that for $Bi\Sigma = 72$, $k_c \to \infty$, thus violates the lubrication approximation $k \ll 1$. Therefore, the proposed model holds good only for $Bi\Sigma < 72$. This event is graphically demonstrated in figure 14. One can readily see that the critical parameters $(Ma_c, k_c)$ predicted by the long-wave theory agree with the numerical results only for $Bi\Sigma = 10(<72)$ (i.e. for a deformable free surface, see panel *a*).

Equations (6.15) and (6.16) indicate that at $k\to\infty$, irrespective of the deformability of the free surface, *Ma* attains the limiting value

$$Ma_{mon.,k\to\infty} = \frac{48Le}{Le + \chi + Le\chi}. \tag{6.18}$$

It should be noted that for $Bi\Sigma \geq 72$, $Ma \to Ma_c$ in equation (6.18). However, this $Ma_c$ lost the quantitative agreement with the numerical results owing to the violation of the long-wave approximation (compare panels (*a,b*) in figure 14).

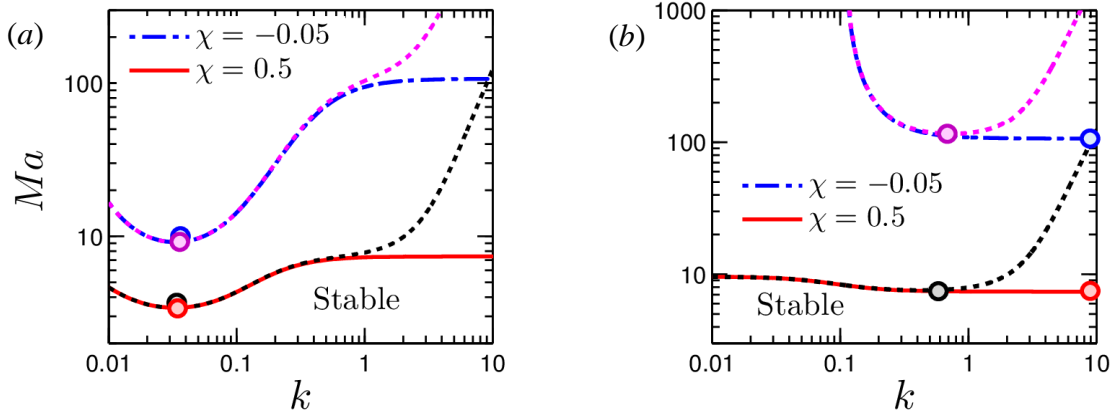

FIGURE 14. (Color online) Comparison of the results obtained from the approximate model (solid and the dash-dotted lines) with the numerical results (dotted lines) for the monotonic instability mode via the neutral stability curve. (*a*) deformable surface $(Ga,\Sigma) = (0.1,10^3)$, (*b*) non-deformable surface $(Ga,\Sigma) \to \infty$. The dot (o) mark on each neutral curve represents the critical point of the curve. Other parameters: $Bi = 0.01$, $Le = 0.1$.

From (6.15)−(6.16) it further follows that, for $k \in [0,\infty)$, irrespective of the deformability of the free surface, $Ma$ remains positive for $\chi > 0$ and becomes negative when $\chi < -Le/(1+Le)$. Therefore, the neutral curves must become discontinuous in the domain $0 > \chi > -Le/(1+Le)$, suggesting the emergence of instability for heating from the top as well as from the bottom. For the case of a deformable free surface, this point of discontinuity ($k_d$) is given by the real and positive root of

$$\gamma_0 k_d^4 + \gamma_1 k_d^2 + \gamma_2 = 0, \tag{6.19}$$

the coefficients $\gamma_j$ ($j = 0-2$) are presented in appendix B.

For a non-deformable surface, the point of discontinuity ($k_d^{nd}$) lies at

$$k_d^{nd} = \sqrt{-\frac{Bi\chi}{Le + \chi + Le\chi}}, \tag{6.20}$$

Thus, for $\chi \in (0, -Le/1+Le)$, instability sets in for heating the liquid layer from below with $k > k_d$ (or $k_d^{nd}$ for a non-deformable surface) and vice versa. For the parameter values $(Le, \chi)$ $=(0.1,-0.05)$ this situation is graphically presented in figure 14. Here, $k_d = 2.8 \times 10^{-3}$ and $k_d^{nd} = 0.105$ (see panels *a,b*). A comprehensive study of the stability characteristics of the branch pertaining to the negative values of *Ma* is beyond the scope of the present paper.

Finally, a few previously reported results in the literature can be derived from the expressions for the neutral curves (6.15) and (6.16). Since the monotonic instability threshold remains unaffected by the elastic behaviour of the fluid; therefore in the limit $\chi = 0$, equation (6.15) yields the results for purely thermocapillary induced convection in a Newtonian liquid layer (Shklyaev, Alabuzhev & Khenner 2012). Furthermore, for such a film, in the limit $(Ga, \Sigma) \to \infty$ equation (6.16) yields the well-known asymptotic $Ma_{mon.} = 48$ either for $Bi = 0$ or $k \to \infty$ (Pearson 1958).

### 6.2. *Oscillatory mode*

The stability threshold for this instability mode is obtained by substituting $\lambda_o = i\omega$ in (6.13). This yield

$$Ma_{\pm osc.} = \frac{(\mathcal{N}_0 - \omega_\pm^2 \mathcal{N}_2)(\mathcal{D}_0 - \omega_\pm^2 \mathcal{D}_2) + \omega_\pm^2 (\mathcal{N}_1 - \omega_\pm^2 \mathcal{N}_3)(\mathcal{D}_1 - \omega_\pm^2 \mathcal{D}_3)}{(\mathcal{D}_0 - \omega_\pm^2 \mathcal{D}_2)^2 + \omega_\pm^2 (\mathcal{D}_1 - \omega_\pm^2 \mathcal{D}_3)^2}, \qquad (6.21)$$

where $\omega_\pm = (\omega_+, \omega_-)$ are the oscillation frequencies for $Ma_{\pm osc.} = (Ma_{+osc.}, Ma_{-osc.})$ respectively and given by

$$\omega_\pm^2 = \frac{1}{2(\mathcal{N}_3 \mathcal{D}_2 - \mathcal{N}_2 \mathcal{D}_3)} \Big[ (\mathcal{N}_1 \mathcal{D}_2 + \mathcal{N}_3 \mathcal{D}_0 - \mathcal{N}_2 \mathcal{D}_1 - \mathcal{N}_0 \mathcal{D}_3) \\ \pm \sqrt{(\mathcal{N}_2 \mathcal{D}_1 + \mathcal{N}_0 \mathcal{D}_3 - \mathcal{N}_1 \mathcal{D}_2 - \mathcal{N}_3 \mathcal{D}_0)^2 - 4(\mathcal{N}_3 \mathcal{D}_2 - \mathcal{N}_2 \mathcal{D}_3)(\mathcal{N}_1 \mathcal{D}_0 - \mathcal{N}_2 \mathcal{D}_3)} \Big]. \qquad (6.22)$$

Note that $\omega_+^2$ ($\omega_-^2$) refers to the oscillation frequency obtained from adding (subtracting) the square root terms in the numerator of (6.22). For a given set of model parameters $(\mathcal{B}, \mathcal{C}a, \mathcal{D}e, Ga, Le, \chi)$, the presence of two different oscillation frequencies (i.e. $\omega_+$ and $\omega_-$) suggests the possible emergence of two different oscillatory instabilities in the system (namely, oscillatory−I and oscillatory−II as discussed in § 4). Confirming the numerical results, one of the oscillation frequencies vanishes for a non-deformable surface (must be the one related with the oscillatory−II mode), for which the expression for the neutral curve and the oscillation frequency is given by

$$Ma_{osc.} = \frac{48(Bi + k^2 + Lek^2)}{\left[(1 + DeLek^2) + \chi(1 + BiDe + Dek^2 + DeLek^2)\right]k^2}, \qquad (6.23)$$

and

$$\omega^2 = \frac{\mathcal{L}_0 k^6 + \mathcal{L}_1 k^4 + \mathcal{L}_2 k^2 + \mathcal{L}_3}{\mathcal{L}_4 k^2 + \mathcal{L}_5}. \qquad (6.24)$$

respectively. The coefficients $\mathcal{L}_j$ ($j = 0-5$) are defined in appendix C. It is worth noting that in the limit $De = 0$, equation (6.23) matches with the results for a Newtonian fluid (Podolny *et al.* 2005).

Given the convoluted form of the neutral curves (6.21) and (6.23), any further analytical progress in estimating their validity bound has now become a rather intricate task. Thus, we verify the accuracy of this analysis by comparing the results with the numerically calculated results for a wide range of the parameters $(De, Le, \chi)$ within the limit $Bi\Sigma < 72$. For the long-wave branch of the oscillatory-I mode, figure 15(*a*) shows that the results obtained from the approximate model agree with the numerical results in an excellent manner for both the Newtonian ($De = 0$) and highly viscoelastic ($De = 10$) binary mixture. Here, the oscillation frequency is given by $\omega_-$. However, the agreement between the results obtained from these two different approaches is found to be poor for the short-wave branch (due to the violation of the lubrication approximation, see figure 15*b*). Nevertheless, the scaling adopted for *De* (see 6.2*b*) helps in achieving a qualitative agreement for higher values of *De*. The oscillation frequency for this short-wave branch is given by $\omega_+$.

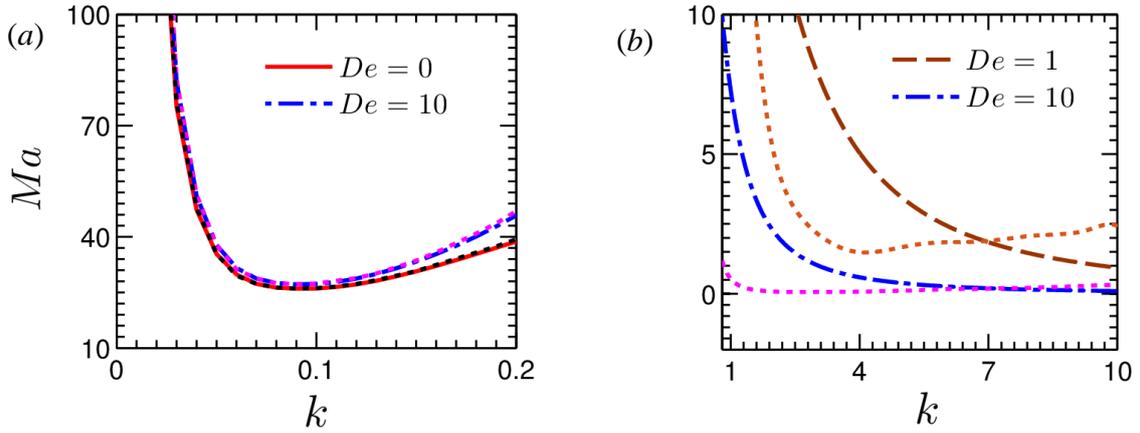

FIGURE 15. (Color online) Comparison between the results of numerical calculation (dotted lines) and the approximate model (solid and the dash-dotted lines) for the oscillatory-I mode via the neutral stability curve: (a) long-wave branch at $Le = 0.1$, $\chi = -0.5$ (b) short-wave branch at $Le = 10^{-3}$, $\chi = 0.5$. Other parameters: $Bi = 0.01$, $Ga = 0.1$, $\Sigma = 10^3$.

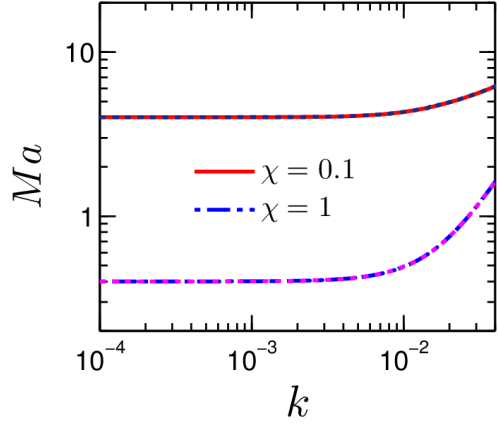

FIGURE 16. (Color online) Comparison of results obtained from the approximate model (solid and the dash-dotted lines) with the numerical results (dotted lines) for the oscillatory-II mode via the neutral stability curve at $Bi = 0.01$, $De = 1$, $Le = 0.1$, $Ga = 0.1$, $\Sigma = 10^3$.

Finally, this approximate model also predicts the appearance of the oscillatory-II instability in the system. The oscillation frequency for this mode is given by $\omega_-$. Figure 16 shows that the developed model is capable of depicting the stability threshold in a fairly accurate manner for this instability mode as well.

## 7. Phase diagrams

We have now understood that both monotonic and oscillatory instabilities can emerge in the present system (either in the long-wave or short-wave mode) depending on the values of the model parameters. The purpose of this section is to explore the parameter regions for each instability mode, wherein it becomes dominant in the fluid layer. The phase diagrams displayed in figure 17 are expected to be helpful for carrying out an experimental investigation of the present problem, especially in situations where one is interested in observing the convective patterns of a particular instability mode. Note that, since we are studying here the stability characteristics of a viscoelastic film incorporating the Soret effect, the phase diagrams are plotted in the $\chi - De$ plane. In an effort to identifying the region of dominance for each instability mode, the parameter set $(Ga, \Sigma, Le)$ is varied to take into account the surface deformability with weak/strong solute diffusivity. However, we hold the parameters $Bi (= 0.01)$ and $Pr (= 10)$ fixed. In panels (*a-d*) of figure 17, region-1 stands for the monotonic mode, region-2 for the long-wave oscillatory-I mode, region 3 for the short-wave oscillatory-I mode, and region-4 for the oscillatory-II mode. An important point that needs to be highlighted here is that a dataset in $(\chi, De)$ corresponding to the boundary between the adjacent instability modes refers to a competition between them to become the dominant instability mode in the system. Such interaction between the instability modes may lead to the formation of convective patterns, which will be significantly different from the patterns that appear far from this location.

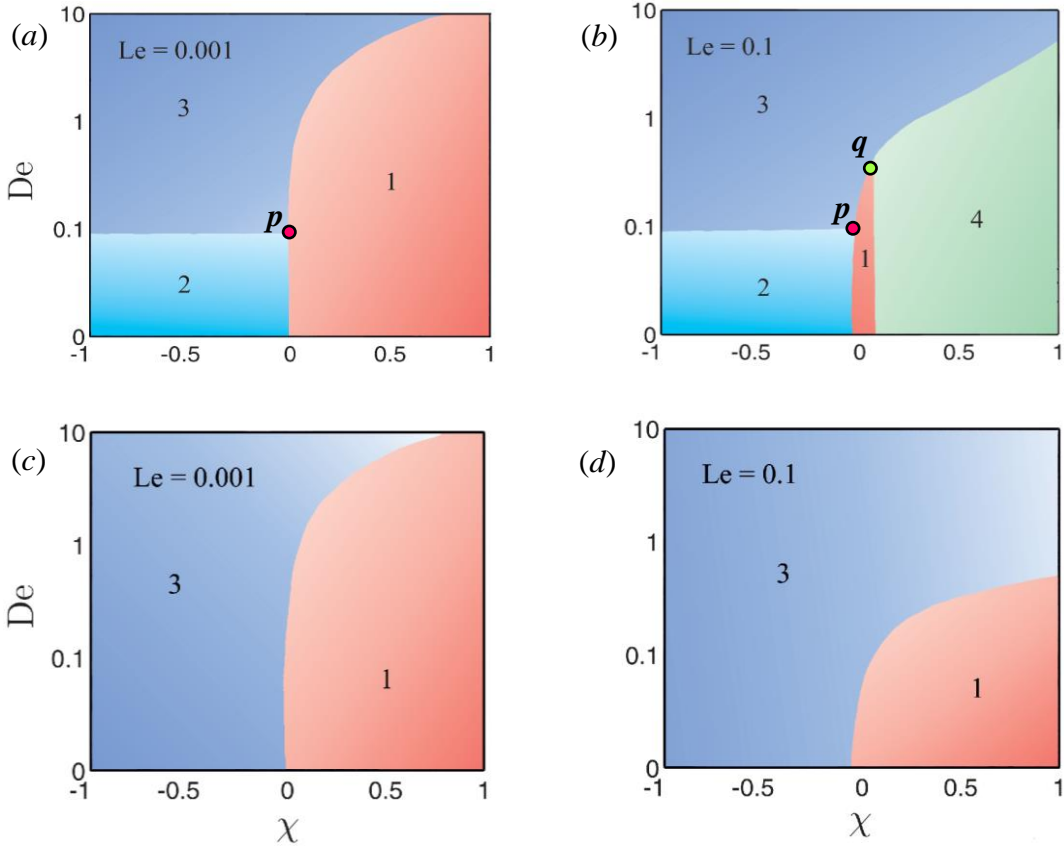

FIGURE 17. (Color online) Phase diagrams for $(\chi, De)$ at different $Le$ enclosing the regimes of dominant instability mode: ($a,b$) deformable free surface $(Ga, \Sigma) = (0.1, 10^3)$, and ($c,d$) non-deformable free surface $(Ga, \Sigma) \to \infty$ at $Bi = 0.01$. In panels ($a$-$d$), regime-1: monotonic instability, regime-2: long-wave oscillatory–I instability, regime-3: short-wave oscillatory–I instability, regime-4: oscillatory –II instability. At the points marked $p$ and $q$ in panels ($a,b$), the three adjacent instability modes can coexist.

For a liquid layer with a deformable free surface, panels ($a,b$) plot the phase diagrams for two different values of $Le$: $Le = 10^{-3}$ and 0.1, respectively. Panel ($a$) shows that in the weakly viscoelastic regime (i.e. $De \lesssim 0.1$), the long-wave oscillatory-I mode (region 2) becomes dominant in the system for $\chi < 0$. The characteristics of this mode are identical for both the Newtonian and viscoelastic binary mixtures, as we have learned from § 4.1. However, in the highly elastic regime (i.e. for $De > 0.1$), this long-wave mode is replaced by its short-wave counterpart (region-3). Since the onset of this particular instability mode is triggered by the enhanced elasticity of the mixture (but weakly dominated by the solutocapillary force), the short-wave oscillatory-I mode can become the dominant instability mode even for $\chi > 0$ at higher values of *De*. Except for such larger values of *De*, the monotonic instability (region-1) prevails in the system for $\chi > 0$ as well as for a narrow interval of $\chi < 0$ (as discussed in §6.1).

A comparison between panels (*a*) and (*b*) now reveals that for a viscoelastic mixture with higher solute diffusivity $(Le = 0.1)$, the region of dominant monotonic instability shrinks drastically, and the oscillatory-II instability (region-4) emerges in the system for $\chi > 0$. Furthermore, region-1 shifts slightly towards left (due to widening up the range $\chi \in (0, -Le/Le+1)$, see § 6.1), and region-3 expands towards the right. However, the transition boundary between the long-wave oscillatory-I (region-2) and short-wave oscillatory-I (region-2) mode remains unaffected by the value of *Le*. A key observation from panels (*a,b*) is that, for $(\chi, De)$ values corresponding to the points ***p*** and ***q***, three different instability modes, *viz.*, monotonic:long-wave oscillatory-I:short-wave oscillatory-I and monotonic:long-wave oscillatory-II:short-wave oscillatory-I respectively, can compete together in the system.

For a non-deformable free surface, panels (*c,d*) demonstrate that irrespective of the diffusivity ratio *Le*, the conductive state bifurcates either to the monotonic (region-1) or the short-wave oscillatory-I (region-3) mode depending upon the parameter set $(\chi, De)$. It should be noted that the long-wave oscillatory mode (both oscillatory-I and oscillatory-II) cannot become the dominant instability mode here (due to the dampening out of such perturbations by the increased gravitational and surface tension forces).

Thus, the panels (*a-d*) provide a comprehensive review of the stability picture under the parameter space $(De, \chi, Le, Ga, \Sigma)$. Besides exploring the parameter regions for which a particular instability mode becomes dominant in the system, they also highlight the competition between the various instability modes that may occur based upon a few specific parameters set. We expect these phase diagrams will provide valuable guidance in choosing the parameters set for a realistic experimental setup.

## 8. Summary and Conclusions

In this work, we have investigated the Marangoni instability problem for a thin viscoelastic film confined between its deformable free surface and a flat rigid substrate. The novelty of this analysis lies at considering the binary aspect of the fluid, along with the incorporation of the Soret effect. Linear stability analysis performed around a quiescent basic state reveals that thermosolutocapillarity, in the presence of the Soret diffusion, shows an entirely different stability picture than from the case of purely thermocapillary driven convection (Sarma & Mondal 2019). For the system subjected to heating from below, it is found that apart from the monotonic instability, two different oscillatory instabilities, namely the oscillatory-I and the oscillatory-II, can emerge in the system depending on the physical situations governed by the parameter space $(Bi, De, Le, \chi, Ga, \Sigma)$ (see figure 17).

The monotonic instability threshold remains unaffected by the elastic behaviour of the mixture, thus resulting in identical stability characteristics for both the Newtonian and viscoelastic liquid films. However, the instability mode, which attracts particular attention in the case of Marangoni convection in a viscoelastic film is the oscillatory instability mode. The oscillatory-I instability may appear either in the long-wave or short-wavelength form depending on the deformability of the free surface and the elasticity level of the fluid. It is found that while the short-wave oscillatory-I mode is more universal, the long-wave oscillatory-I mode can also get dominant in a weakly viscoelastic film ($De \lesssim 0.1$) with a deformable free but exclusively for $\chi < 0$. The oscillatory-II mode is even more case-specific and likely to emerge in a binary liquid film (irrespective of the elasticity of the mixture) having a deformable free surface with high solute diffusivity $Le \gtrsim O(10^{-2})$ and only for $\chi > 0$. This is a long-wavelength instability that appears with a significantly large oscillation period compared to the oscillatory-I mode.

This leads to the following main conclusions from this study: the solutocapillary effect plays a crucial role only in the case of long-wave disturbances. For $\chi > 0$ (or equivalently $\mathcal{S} > 0$), it causes the emergence of long-wave instability in terms of the monotonic or the oscillatory-II mode depending on the solute diffusivity. For $\chi < 0$, solutocapillarity enhances the stability of the system against the long-wave oscillatory-I perturbations. On the other hand, the thermocapillary effect is primarily responsible for the short-wave oscillatory-I disturbances. The solutocapillarity plays here a minor role. Triggered by the elasticity of the mixture, the short-wave oscillatory-I instability can, therefore, appear for any $\chi \in \mathbb{R}$. While the increased (convective) heat transfer rate and reduced deformability of the free surface enhance the stability of the system against the long-wave perturbations, the increasing elasticity of the fluid makes the system more vulnerable towards the short-wave disturbances.

However, the conditions (*viz.* the film thickness, the temperature difference across the film, the corresponding size of the convection cell and its oscillation period) at which one may experimentally observe a particular instability mode detected in this study remains here unclear. This is primarily due to uncertainty over data related to the physical properties of the fluid, especially the Soret coefficient $\mathcal{S}$ and the relaxation time $\lambda$ that changes with the composition of the fluid. Separate experimentation is needed for this purpose. Nevertheless, provided a prior estimation of the parameters $(\lambda, \mathcal{S})$, the present analysis can be helpful in predicting the instability modes as soon as a bifurcation of the conductive base state occurs. Further experimental and theoretical investigations are thus necessary to understand the pattern dynamics in the post-critical regime for the present convection phenomenon. It is also worth extending the present model to study an unsteady problem e.g. the evaporative Marangoni convection (Doumenc *et al.* 2010; Pillai & Narayanan 2018) in a polymeric film, or to develop a coupled film-substrate model (Batson *et al.* 2019) to analyse the convection phenomenon for more complex systems.

## Acknowledgements

The authors are grateful to the referees for their valuable comments and suggestions that significantly improved the manuscript. They would also like to acknowledge the computing facilities available at the Microfluidics and Microscale transport process laboratory, IIT Guwahati, where this work was carried out.

## Appendix A. Expressions for the coefficients used in (6.13)

The coefficients of (6.13) are as follows

$$\mathcal{N}_0 = 2\mathcal{B}\mathcal{Q}q^4 + 2\mathcal{Q}q^6, \tag{A.1}$$

$$\mathcal{N}_1 = -2\mathcal{B}\left(3 + \mathcal{Q}Le^{-1}\right)q^2 - 2\left[3 + \mathcal{Q}\left(1 + Le^{-1} + \mathcal{B}\mathcal{D}e\right)\right]q^4 - 2\mathcal{D}e\mathcal{Q}q^6, \tag{A.2}$$

$$\mathcal{N}_2 = 6\mathcal{B}Le^{-1} + 2\left[\left(3 + \mathcal{Q} + \mathcal{B}\mathcal{D}e\mathcal{Q}\right)Le^{-1} + 3\right]q^2 + 2\mathcal{D}e\mathcal{Q}\left(1 + Le^{-1}\right)q^4, \tag{A.3}$$

$$\mathcal{N}_3 = -6Le^{-1} - 2\mathcal{D}e\mathcal{Q}Le^{-1}q^2; \tag{A.4}$$

$$\mathcal{D}_0 = \mathcal{B}\mathcal{Q}\chi Le^{-1}q^2 + \left[(1+\chi)(72+\mathcal{Q}) + \mathcal{Q}\chi Le^{-1}\right]q^4, \tag{A.5}$$

$$\begin{aligned}\mathcal{D}_1 &= -48\mathcal{B}\chi Le^{-1} - \left[(1+\chi)\left(48 + \mathcal{Q}Le^{-1}\right) + 2\left(36 + 24\chi + \mathcal{D}e\chi\mathcal{Q}\right)Le^{-1}\right]q^2 \\ &\quad - 2\mathcal{D}e\left[\mathcal{Q}\chi Le^{-1} + (1+\chi)(36+\mathcal{Q})\right]q^4\end{aligned}, \tag{A.6}$$

$$\begin{aligned}\mathcal{D}_2 =\ & 48(1+\chi+\mathcal{B}\mathcal{D}e\chi)Le^{-1}+\mathcal{D}e^2\mathcal{Q}(1+\chi+\chi Le^{-1})q^4 \\ & +\mathcal{D}e\left[2(1+\chi)(24+\mathcal{Q}Le^{-1})+(72+48\chi+\mathcal{B}\mathcal{D}e\chi\mathcal{Q})Le^{-1}\right]q^2,\end{aligned} \quad (A.7)$$

$$\mathcal{D}_3 = -\mathcal{D}e(1+\chi)(48+\mathcal{D}e\mathcal{Q}q^2)Le^{-1}. \quad (A.8)$$

Recall $\mathcal{Q} = Ga + Ca\,q^2$.

## Appendix B. Expressions for the coefficients utilized in (6.19)

The expressions for the coefficients of (6.19) are given below

$$\gamma_0 = \Sigma(Le+\chi+Le\chi), \quad (B.1)$$

$$\gamma_1 = \chi(Ga+Bi\Sigma)+Le(1+\chi)(72+Ga), \quad (B.2)$$

$$\gamma_2 = BiGa\chi. \quad (B.3)$$

## Appendix C. Expressions for the coefficients used in (6.21)

The coefficients of (6.21) are

$$\mathcal{L}_0 = DeLe(Le+\chi+Le\chi), \quad (C.1)$$

$$\mathcal{L}_1 = Le^2(1+\chi)(BiDe-1)+2BiDeLe\chi-\chi(1+Le), \quad (C.2)$$

$$\mathcal{L}_2 = Bi\chi(BiDeLe-Le-2), \quad (C.3)$$

$$\mathcal{L}_3 = -Bi^2\chi, \quad (C.4)$$

$$\mathcal{L}_4 = -De, \quad (C.5)$$

$$\mathcal{L}_5 = 1+\chi-BiDe. \quad (C.6)$$